\documentclass[a4paper,times]{cjeart}

\usepackage{amsthm} %

\newtheorem{corollary}{Corollary}
\newtheorem{lemma}{Lemma}

\usepackage{amsmath,amssymb,bm}
\usepackage{algorithm}
\usepackage{algpseudocode}
\usepackage{subcaption}

\usepackage{framed}
\usepackage{algpseudocode}
\usepackage{bm}
\usepackage{stfloats}  
\usepackage{setspace}
\usepackage{balance}

\usepackage{amsmath,amssymb,amsfonts}
\allowdisplaybreaks[3]
\usepackage{caption} 
\usepackage{xcolor}
\usepackage[T1]{fontenc}   

\makeatletter
\renewcommand{\maketag@@@}[1]{\hbox{\m@th\normalsize\normalfont#1}}%
\makeatother

\renewcommand{\Re}{\operatorname{Re}}

\makeatletter

\newcommand{\Rmnum}[1]{\expandafter\@slowromancap\romannumeral #1@}

\makeatother

\graphicspath{{figures/}}

\volumeyear{}
\volumenumber{}
\issuenumber{}
\journalname{Chinese Journal of Electronics}
\startpage{1}
\DOI{}
\journaltype{Research Article}

\title[Joint Transceiver Orientation Optimization for Rotatable-Antenna MIMO Capacity Maximization]{Joint Transceiver Orientation Optimization for Rotatable-Antenna MIMO Capacity Maximization}

\author{%
	Ailing Zheng\affilnums{1}, 
	Qingqing Wu \affilnums{1}, 
	Xingxiang Peng \affilnums{1}, 
	Qiaoyan Peng \affilnums{1,2}, 
	Ziyuan Zheng \affilnums{1},  and \\
	Wen Chen \affilnums{1}
}

\affiliation{%
	\affilnum{1} School of Information Science and Electronic Engineering, Shanghai Jiao Tong University, Shanghai 200240, China \\
	\affilnum{2} State Key Laboratory of Internet of Things for Smart City and the Department of Electrical and Computer Engineering, University of Macau, Macao SAR, China
}

\corrauth{Qingqing Wu}
\email{qingqingwu@sjtu.edu.cn}

\receivetime{Received April 30, 2026}
\accepttime{}
\publishtime{}

\authorcopy{Ailing Zheng\emph{et al.}}

\abstract{%
	Conventional multiple-input multiple-output (MIMO) systems mainly rely on fixed antenna arrays, which limits their capability to adapt the effective channel matrix to the propagation environment. Rotatable antennas (RAs), which enable mechanical or electronic adjustment of antenna boresight directions, introduce a new orientation-domain degree of freedom for channel reconfiguration. In this paper, we investigate an RA-aided MIMO communication system for channel capacity enhancement. We first establish an orientation-dependent MIMO channel model. Then, we formulate a capacity maximization problem by jointly optimizing the transmit covariance matrix and the transmit/receive RA orientations under practical spherical-cap constraints. To solve this non-convex problem, we develop an alternating optimization algorithm, where the transmit covariance matrix is updated via eigenmode transmission and water-filling, while each RA orientation is optimized through a Riemannian Frank-Wolfe method. We further investigate the low-SNR regime and derive simplified designs for multiple-input single-output (MISO) and single-input multiple-output (SIMO) special cases. Numerical results show that the proposed RA-aided MIMO design significantly improves the channel capacity compared with the fixed-orientation benchmark, demonstrating the benefits of orientation-domain channel reconfiguration.
}


\begin{document}
	\maketitle	
	\vspace{-2mm}
	\section{Introduction}
	The rapid proliferation of emerging wireless applications, such as extended reality (XR), intelligent transportation, and immersive mobile services, has imposed increasingly stringent requirements on the capacity, spectral efficiency, and reliability of future wireless communication systems \cite{Zhang20196G}. Multiple-input multiple-output (MIMO) communication has been widely recognized as one of the most important enabling technologies for improving wireless throughput by exploiting spatial multiplexing and array gains \cite{Saad6G,Chowdhury6G2020}. By deploying multiple antennas at both the transmitter and the receiver, MIMO systems are capable of supporting parallel data-stream transmission over multiple spatial eigenchannels, thereby significantly enhancing the channel capacity compared with single-antenna systems \cite{LuMIMO2014}. However, conventional MIMO systems usually employ fixed antenna arrays. Once deployed, the physical locations and boresight directions of antenna elements are generally predetermined and cannot be flexibly adjusted according to the instantaneous propagation environment \cite{InterdonatoMIMO2020}. As a result, the effective channel matrix is mainly determined by the surrounding scattering environment and the fixed array geometry. Although conventional beamforming and precoding techniques can adapt the transmitted signals in the digital or analog domain, they cannot directly reconfigure the radiation direction of each antenna element. This limits the ability of conventional MIMO systems to fully exploit the spatial and angular variations of wireless channels, especially when the dominant line-of-sight (LoS) and non-LoS (NLoS) paths arrive from different spatial directions.
	
	To further enhance MIMO performance, various advanced antenna and array architectures have been investigated. Massive MIMO improves the spectral efficiency by deploying a large number of antennas, but it generally requires more radio-frequency chains, higher hardware cost, and increased power consumption \cite{Teng2022}. Hybrid analog-digital beamforming has been widely studied for millimeter-wave systems to reduce hardware complexity, yet its performance is constrained by the limited flexibility of analog phase shifters and fixed array structures \cite{Kaushik2019}.
	To overcome the above limitations, movable antenna (MA) \cite{Zhu2024MA} and fluid antenna system (FAS) \cite{Wong2021} have been proposed, where the antenna positions can be flexibly adjusted in the given region to reshape the channel characteristics. Thus, the MIMO channel matrix between them can be reshaped for higher capacity. Specifically, the authors of \cite{Ma2024MIMO} investigated the capacity of MA-enabled MIMO communication systems by optimizing the positions of transmit and receive MAs as well as the covariance of transmit signals. 
	By relocating antennas to favorable positions and reconfiguring the array geometry, the system enhances desired signal strength, suppresses interference, and supports adaptive null steering \cite{Ma2024MIMO,New2024,Zhu2024CM}. Such flexible rearrangement also reshapes the channel matrix to improve spatial multiplexing, thereby introducing additional spatial degree of freedoms (DoFs) and intelligently reshaping the wireless channel to improve overall communication performance \cite{Zhu2024CM}. 
	Owing to these advantages, MAs have been incorporated into various emerging wireless applications, including integrated sensing and communications
	(ISAC) \cite{Guo2024MA,Li2025,Wang2025MA-IRS}, unmanned aerial vehicle communications \cite{Liu2025,Tang2025}, and non-orthogonal multiple access systems \cite{Xiao2025NOMA,Gao2025,Zhou2024NOMA}. 
	
	However, despite their remarkable performance gains, MAs may face practical limitations in fast-varying channels due to finite mechanical response time and movement speed. Moreover, conventional MA architectures mainly exploit position-domain reconfigurability, while the antenna orientations are usually kept fixed. This restricts their capability to fully adapt to dynamic propagation environments. To address these drawbacks, the six-dimensional (6D) MA architecture was proposed in \cite{Shao2025-1}, where both the position and orientation of each antenna surface can be jointly adjusted according to channel variations. The effectiveness of 6DMA has been verified in enhancing network capacity \cite{Shao2025-2}, improving physical-layer security \cite{Qiansecure2026}, and strengthening spatial multiplexing performance \cite{Ren2025}.
	As a lightweight version of 6DMA, rotatable antennas (RAs) have recently attracted increasing attention due to their reduced hardware complexity and compact implementation. By dynamically adjusting the three-dimensional (3D) boresight direction of each antenna element, RA-enabled systems can reconfigure the angular-domain channel gains and thus enhance the overall array performance \cite{Peng2025,Zheng2026}. Compared with conventional fixed-orientation antennas, RAs introduce additional orientation-domain DoFs, enabling the radiated energy to be directionally concentrated and physically steered toward desired spatial directions. This capability can further improve the effective array gain for intended users.
	Motivated by these advantages, RA-enabled wireless communication systems have recently been studied in various scenarios, including ISAC \cite{Zhou2025,Zheng2025MA-4}, secure communication \cite{Liang2025,Jiang2025RA}, and unmanned aerial vehicle systems~\cite{Zhang2025RA,ChenUAV2026}. 
	
	Although RA-enabled wireless systems have demonstrated promising
	performance gains in various scenarios, the fundamental capacity
	characterization of RA-aided MIMO systems remains largely unexplored.
	Different from single-stream transmission, the performance of MIMO
	systems depends not only on the effective channel power, but also on the
	eigenmode structure of the channel matrix, which determines the achievable
	spatial multiplexing gain. In RA-aided MIMO systems, antenna rotations can
	modify the directional gains of different propagation paths and thereby
	reshape the effective channel matrix. As a result, RA orientation design
	is intrinsically coupled with transmit covariance optimization, which
	motivates a systematic study of capacity-oriented joint transmit and
	orientation design.
	
	In this paper, we investigate an RA-aided MIMO communication system,
	where both the transmitter and the receiver are equipped with uniform
	planar arrays (UPAs) composed of RAs. Different from conventional
	fixed-orientation arrays, the channel coefficients in the considered
	system depend on the propagation distances, propagation phases, and
	transmit/receive antenna orientations through the directional antenna
	gain pattern. In many practical scenarios, the array panel is
	mechanically fixed after installation, and changing the overall array
	pose may be restricted by installation space, device form factor, and
	stability requirements.
	Therefore, this paper focuses on element-wise RAs, where each antenna
	element adjusts its boresight direction while the array position and panel
	pose remain fixed.
	This element-wise orientation control provides an additional DoF for
	reconfiguring the effective MIMO channel without changing the array
	geometry.
	However, the additional orientation-domain flexibility also makes the
	system design more challenging. For RA-aided MIMO systems, capacity
	maximization requires the joint optimization of the transmit covariance
	matrix and all transmit/receive RA orientations. The transmit covariance
	matrix determines the power allocation among spatial eigenmodes, while
	the RA orientations affect the singular values and eigenstructure of the
	effective MIMO channel. Therefore, efficient optimization methods are
	needed to exploit the orientation-domain channel reconfiguration
	capability under practical rotation constraints.
	
	Motivated by the above considerations, this paper studies the joint
	transmit design and RA orientation optimization for RA-aided MIMO
	communication systems. We characterize the MIMO channel capacity by
	jointly optimizing the transmit covariance matrix and the transmit/receive
	RA orientations under practical spherical-cap constraints. The main
	contributions of this paper are summarized as follows.

	\begin{itemize}
		\item We establish a general RA-aided MIMO channel model that incorporates both LoS and NLoS propagation components. Based on this model, we study a RA-aided MIMO system and formulate a capacity maximization problem via the joint design of the transmit covariance matrix and the orientations of all transmit and receive RAs. The resulting problem is highly non-convex due to the coupling between the transmit covariance matrix and the orientation-dependent channel matrix, as well as the zenith-angle constraints imposed on each RA orientation.
		
		\item To address the formulated problem, we propose an alternating optimization (AO) framework. In particular, the optimal transmit covariance matrix is obtained through eigenmode transmission and water-filling. The orientation of each RA is optimized separately, which decomposes the original high-dimensional problem into a sequence of more tractable subproblems. For each orientation update, we further design an efficient Riemannian Frank-Wolfe algorithm over the spherical-cap feasible set.
		
		\item We further investigate several important special cases and asymptotic regimes. In the low-SNR regime, the original problem is shown to reduce to dominant eigenchannel enhancement. Moreover, for the multiple-input single-output (MISO) and single-input multiple-output (SIMO) cases, the capacity maximization problem is simplified to channel power maximization, which provides useful insights and simplified objective functions for
		orientation optimization.
		
		\item Numerical results verify the effectiveness of the proposed RA-aided MIMO design. By exploiting the additional orientation-domain DoFs, the proposed algorithm significantly improves the MIMO channel capacity compared with conventional fixed-orientation benchmark schemes. In particular, the proposed scheme achieves up to a $34\%$ performance gain over the fixed-antenna baseline.
	\end{itemize}
	
	The rest of this paper is organized as follows. Section~\ref{System Model} presents the system model and formulates the RA-aided MIMO capacity maximization problem. Section~\ref{Proposed Algorithm} develops the AO algorithm. Section \ref{Simulation Results} provides numerical results to validate the performance of the proposed algorithm. Finally, Section \ref{Conclusion} concludes this paper.
	
	\emph{Notations:} 
	$\mathrm{Tr}(\mathbf{A})$ and $\text{det} (\mathbf{A}) $ denote the trace operation and the determinant of matrix $\mathbf{A}$. $\mathbf{A}^{\dagger}$ represents the pseudo-inverse of matrix $\mathbf{A}$. For a square matrix $\mathbf{A}$, $\mathbf{A}^{-1}$ denotes its inverse if it is nonsingular. $\nabla_{\mathbf{x}} f(\mathbf{x})$ represents the gradient vector of the function $f$ w.r.t. the vector $\mathbf{x}$. $\nabla$ denotes the differential operator. $\mathcal{CN} (\mathbf{0},\mathbf{Z})$ denotes the circularly symmetric complex Gaussian distribution with zero mean and covariance matrix $\mathbf{Z}$. 
	$\mathbb{C}$ and $\mathbb{R}$ denote complex field and real field, respectively. $[x]_{+}=\max\{0,x\}$. $\mathbf{I}_M$ denotes the $M$-dimensional identity matrix.

	\section{System Model}
	\label{System Model}
	\begin{figure}[t]
		\centering
		\includegraphics[width=0.45 \textwidth]{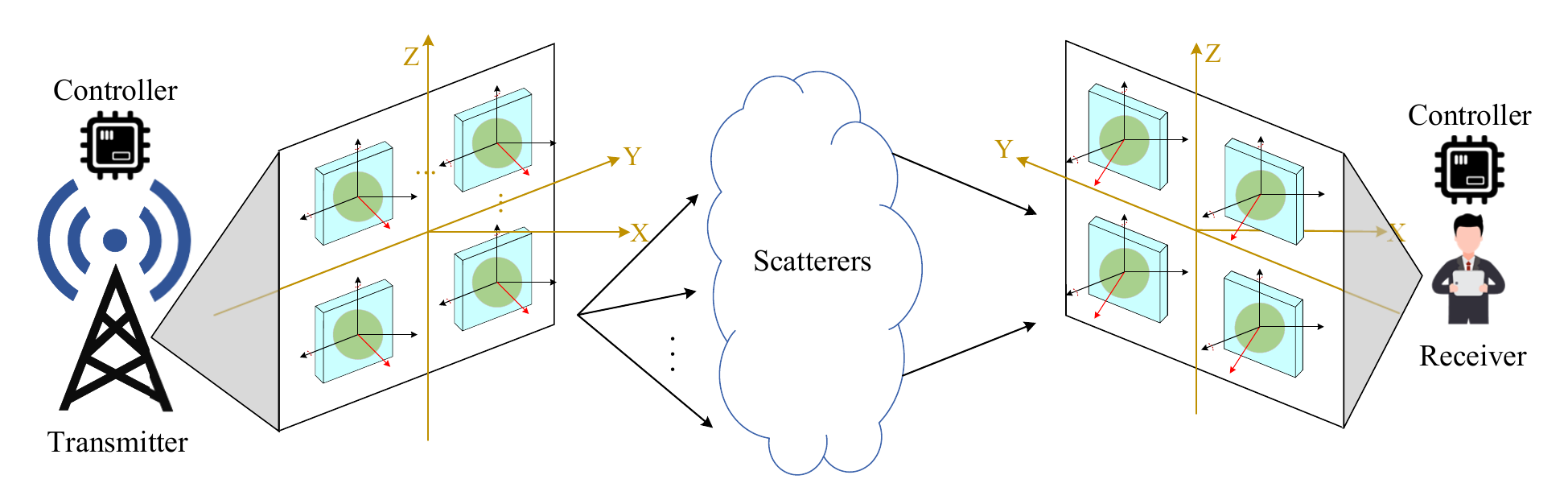}
		\caption{An RA-aided MIMO system, where both the transmitter and
			receiver are equipped with UPAs composed of RAs.}
		\label{SystemModel}
		\vspace{-5mm}
	\end{figure}
	We consider a MIMO communication system, as shown in Fig. \ref{SystemModel}. The transmitter is equipped with $N$ transmit RAs and the receiver has $M$ receive RAs, where $N=N_x \times N_y$ and $M=M_x \times M_y$. The UPAs at the transmitter and the receiver are located in the local $x_t-y_t$ plane and $x_r-y_r$ plane, respectively. Their global positions are $\mathbf{t}_0=(x_t,y_t,z_t)$ and $\mathbf{r}_0=(x_r,y_r,z_r)$, respectively. Denote $\mathcal{M}=\{1,2,\ldots, M\}$ and $\mathcal{N}=\{1,2,\ldots, N\}$ as the sets of receive antennas and transmit antennas, respectively.
	The local position of the $n$-th transmit RA and the $m$-th receive RA can be characterized by $\mathbf{t}_{n}^l = [x_{t,n}, y_{t,n}, z_{t,n}]^\mathrm{T}$ and $\mathbf{r}_{m}^l = [x_{r,m}, y_{r,m}, z_{r,m}]^\mathrm{T}$, respectively.
	Let $\mathbf{R}_t$ and $\mathbf{R}_{r}$ denote the rotation matrix from the local coordinate system to the global coordinate system for the transmitter and the receiver, respectively. Define $\mathbf{t}_n$ and $\mathbf{r}_{m}$ as the global position of transmit RA $n$ and receive RA $m$, which is given by
	\setlength\abovedisplayskip{0pt}
	\setlength\belowdisplayskip{1pt}
	\begin{align}
		\mathbf{t}_n = \mathbf{t}_0+\mathbf{R}_t\mathbf{t}_n^l,\mathbf{r}_{m} = \mathbf{r}_0+\mathbf{R}_{r}\mathbf{r}_{m}^l.
	\end{align}
	
	We assume that the position and pose of each UPA remain fixed. Thus, the rotation matrices $\mathbf{R}_t$ and $\mathbf{R}_{r}$ are predetermined during deployment. However, each antenna element on the UPA can independently adjust its orientation through electronic or mechanical means \cite{Peng2025}.
	Denote the 3D boresight direction of transmit RA $n$ in the local coordinate system as
	\begin{align}
		\mathbf{f}_{t,n} = [f_{t,n}^x,f_{t,n}^y,f_{t,n}^z]^T,
	\end{align}
	where $f_{t,n}^x$, $f_{t,n}^y$, and $f_{t,n}^z$ are the projections of transmit RA $n$'s pointing vector on the $x$-, $y$-, and $z$-axes, respectively. Let $\theta_{z,n} $ denote the RA $n$'s zenith angle and $\theta_{a,n}$ represent its azimuth angle in the local coordinate system. Thus, we have $f_{t,n}^x = \sin\theta_{z,n}\cos\theta_{a,n}$, $f_{t,n}^y = \sin\theta_{z,n}\sin\theta_{a,n}$, and $f_{t,n}^z = \cos\theta_{z,n}$. Furthermore, we have $\Vert \mathbf{f}_{t,n} \Vert_2 = 1$ due to normalization. To account for practical rotational constraint and mitigate antenna coupling between any two RAs, the zenith angle of each RA should be confined to a specific range
	\begin{align}
		0 \le \theta_{z,n} \le \theta_{\max},
	\end{align}
	where $\theta_{\max} \in [0, \pi/2]$ is the maximum zenith angle that each RA is allowed to adjust. This angular constraint on $\theta_{z,n}$ is equivalent to the following constraint on the boresight vector
	\begin{align}
		\cos(\theta_{\max}) \le \mathbf{f}_{t,n}^T\mathbf{e}_z \le 1,
	\end{align}
	where $\mathbf{e}_z = [0,0,1]^T$ is the unit vector along the $z$-axis.
	Similarly, the pointing vector of the receive RA $m$ is given by $\mathbf{f}_{r,m}$.
	The corresponding boresight vector in the global coordinate system is expressed as 
	\begin{align}
		\tilde{\mathbf{f}}_{t,n} = \mathbf{R}_{t}\mathbf{f}_{t,n},\tilde{\mathbf{f}}_{r,m} = \mathbf{R}_{r}\mathbf{f}_{r,m}.
	\end{align} 
	Let $\mathbf{f}_{t} \!\!=\!\! [\mathbf{f}_{t,1},\mathbf{f}_{t,2},\ldots,\mathbf{f}_{t,N}]$ and $\mathbf{f}_{r}\!\! =\!\! [\mathbf{f}_{r,1},\mathbf{f}_{r,2},\ldots,\mathbf{f}_{r,M}] $ as the collections of the orientations of $N$ transmit RAs and $M$ receive RAs, respectively. 
	
	The effective antenna gain for each RA depends on the signal arrival/departure angle and antenna directional gain pattern. Let $\epsilon$ denote the angular offset between the signal direction and the antenna's main-lobe boresight. 
	The generic directional gain pattern for each RA is as follows:
	\begin{align}
		\label{antennaGain}
		G(\epsilon) = 
		\begin{cases}
			G_0\cos^{2p}(\epsilon), & \epsilon \in [0, \frac{\pi}{2}], \\
			0, & \text{otherwise},
		\end{cases}
	\end{align}
	where $p \ge 0$ determines the directivity of the antenna, and $G_0$ is the maximum gain in the boresight direction with $G_0=2(2p+1)$ to satisfy the law of power conservation. In particular,
	$p=0$ corresponds to the isotropic-antenna case over the front
	half-space, while $p\ge 1$ corresponds to directional RA patterns.
	In the proposed orientation optimization algorithm, we focus on the
	directional RA case with $p\ge 1$. 
	
	\subsection{Channel Model}
	We consider a narrow-band geometric frequency-flat channel model \cite{Chen2022IRS}. The LoS link power gain from the transmit RA $n$ to the receive antenna $m$ is given by
	\begin{align}
		\!\!\! G_{m,n}^{\mathrm{LoS}}(\mathbf{f}_{t,n},\mathbf{f}_{r,m}) \!=\!&\beta_0r_{m,n}^{-2}G(\epsilon_{m,n})G(\epsilon_{n,m}) \nonumber \\ 
		\!\!\! =&\beta_0r_{m,n}^{-2}G_0^2\left[\!\frac{(\mathbf{R}_t\mathbf{f}_{t,n})^\mathrm{T}(\mathbf{r}_{m}\!-\!{\mathbf{t}}_{n})}{r_{m,n}}\!\right]_{+}^{2p} \nonumber \\ 
		\!\!\!&\left[\!\frac{(\mathbf{R}_r\mathbf{f}_{r,m})^\mathrm{T}(\mathbf{t}_n \!-\!{\mathbf{r}}_{m})}{r_{m,n}}\!\right]_{+}^{2p},
	\end{align}
	where $\beta_0 = (\frac{\lambda}{4\pi })^2$ is the free-space reference gain constant, $\lambda$ denotes the wavelength. The term $r_{m,n} = \Vert {\mathbf{t}_n - \mathbf{r}}_{m}\Vert_2$ is the distance between the transmit RA $n$ and the receive RA $m$, and $\cos(\epsilon_{m,n}) =\frac{(\mathbf{R}_t\mathbf{f}_{t,n})^\mathrm{T}({\mathbf{r}}_{m}-\mathbf{t}_n)}{r_{m,n}}$ represents the cosine of the angle between the transmit RA $n$'s orientation $\mathbf{R}_t\mathbf{f}_{t,n}$ and the LoS direction to the receive RA $m$. Similarly, $\cos(\epsilon_{n,m}) =\frac{(\mathbf{R}_r\mathbf{f}_{r,m})^\mathrm{T}(\mathbf{t}_n-\mathbf{r}_{m})}{r_{m,n}}$ denotes the cosine of the angle between the receive RA $m$'s orientation $\mathbf{R}_{r}\mathbf{f}_{r,m}$  and the LoS direction to the transmit RA $n$. The related channel is expressed as \cite{Peng2025}
	\begin{align}
		&h_{m,n}^{\mathrm{LoS}}(\mathbf{f}_{t,n},\mathbf{f}_{r,m})= \sqrt{G_{m,n}^\mathrm{LoS}(\mathbf{f}_{t,n},\mathbf{f}_{r,m})}e^{-j\frac{2\pi}{\lambda}r_{m,n}} \nonumber \\
		& = \sqrt{\beta_0}r_{m,n}^{-1}G_0{f}_{t,m,n}{f}_{r,m,n}e^{-j\frac{2\pi}{\lambda}r_{m,n}}, \nonumber \\
		& = a_{m,n}{f}_{t,m,n}{f}_{r,m,n},
	\end{align}
	where $a_{m,n}=\frac{\sqrt{\beta_0}G_0}{r_{m,n}}e^{-j\frac{2\pi}{\lambda}r_{m,n}}$ and
	\begin{align}
		&\!\!\!\!\! {{f}}_{t,m,n} \!\!=\!\! \left[\frac{(\mathbf{R}_t\mathbf{f}_{t,n})^\mathrm{T}(\mathbf{r}_{m}-{\mathbf{t}}_{n})}{r_{m,n}}\right]_{+}^{p},   \\
		&\!\!\!\!\!{{f}}_{r,m,n} \!\!=\!\! \left[\frac{(\mathbf{R}_{r}\mathbf{f}_{r,m})^\mathrm{T}(\mathbf{t}_n-{\mathbf{r}}_{m})}{r_{m,n}}\right]_{+}^{p}.
	\end{align}
	We further consider a scattering environment with $D$ spatially distributed clusters, located at $\{\mathbf{s}_d \in \mathbb{R}^3\}^D_{d=1}$ in the global coordinate system. Following the same principle, the NLoS link power gain between transmit RA $n$ and cluster $d$ is
	\begin{align}
		G_{n,d}^{\mathrm{NLoS}}(\mathbf{f}_{t,n}) &=\beta_0r_{n,d}^{-2}G(\epsilon_{n,d}) \nonumber \\
		& =\beta_0r_{n,d}^{-2}G_0\left[\frac{(\mathbf{R}_t\mathbf{f}_{t,n})^\mathrm{T}(\mathbf{s}_d-{\mathbf{t}}_{n})}{r_{n,d}}\right]_{+}^{2p},
	\end{align}
	where $r_{n,d} =\Vert \mathbf{s}_d-{\mathbf{t}}_{n} \Vert_2$ is the antenna-to-cluster distance, and $\cos(\epsilon_{n,d})= \frac{(\mathbf{R}_t\mathbf{f}_{t,n})^\mathrm{T}(\mathbf{s}_d-{\mathbf{t}}_{n})}{r_{n,d}}$ denotes the cosine of the angle between the boresight and the direction to cluster $d$. Similarly, the NLoS link power gain between cluster $d$ and the receive RA $m$ is
	\begin{align}
		\!\!\! G_{d,m}^{\mathrm{NLoS}}(\mathbf{f}_{r,m}) &\!=\!\beta_0r_{d,m}^{-2}G(\epsilon_{d,m}) \nonumber \\
		& \!=\!\beta_0r_{d,m}^{-2}G_0\left[\! \frac{(\mathbf{R}_{r}\mathbf{f}_{r,m})^\mathrm{T}(\mathbf{s}_d \!-\! {\mathbf{r}}_{m})}{r_{d,m}}\! \right]_{+}^{2p}.
	\end{align}
	Considering a bi-static scattering model \cite{Peng2025}, the NLoS channel coefficient from the transmit RA $n$ to the receive RA $m$ is given by
	\begin{align}
		& h_{m,n}^{\mathrm{NLoS}}(\mathbf{f}_{r,m},\mathbf{f}_{t,n}) = \nonumber \\
		&\sum \limits_{d=1}^D \! \sqrt{\frac{\sigma_dG_{n,d}^{\mathrm{NLoS}}(\mathbf{f}_{t,n})G_{d,m}^{\mathrm{NLoS}}(\mathbf{f}_{r,m})}{4\pi }} e^{-j\frac{2\pi}{\lambda}(r_{n,d}+r_{d,m})+j\chi_d}, \nonumber \\
		&\! =\! \sum \limits_{d=1}^D b_{m,n,d}{f}_{t,n,d}{{f}}_{r,d,m},
	\end{align}
	where $b_{m,n,d} = \sqrt{\frac{\sigma_d}{4\pi}} \beta_0r_{n,d}^{-1}r_{d,m}^{-1}G_0e^{-j\frac{2\pi}{\lambda}(r_{n,d}+r_{d,m})+j\chi_d}$, with
	\begin{align}
		&\!\!\!\!\! {f}_{t,n,d} \!\!=\!\! \left[\frac{(\mathbf{R}_t\mathbf{f}_{t,n})^\mathrm{T}(\mathbf{s}_d-{\mathbf{t}}_{n})}{r_{n,d}}\right]_{+}^{p},  \\
		&\!\!\!\!\!{{f}}_{r,d,m} \!\!=\!\! \left[\frac{(\mathbf{R}_{r}\mathbf{f}_{r,m})^\mathrm{T}(\mathbf{s}_d-{\mathbf{r}}_{m})}{r_{d,m}}\right]_{+}^{p}.
	\end{align}
	The term $r_{d,m} = \Vert \mathbf{s}_d - \mathbf{r}_{m}\Vert_2$ is the cluster-to-receive RA $m$ distance, $\sigma_d$ denotes the radar cross section of cluster $d$, and $\chi_d$ is a random phase distributed over $[0, 2\pi)$. Then, we have
	\begin{align}
		h_{m,n} = h_{m,n}^{\mathrm{LoS}}(\mathbf{f}_{r,m},\mathbf{f}_{t,n})+h_{m,n}^{\mathrm{NLoS}}(\mathbf{f}_{r,m},\mathbf{f}_{t,n}). 
	\end{align}
	\begin{align}
		\label{channel_h}
		& \mathbf{H}(\mathbf{f}_{t},\mathbf{f}_{r}) \!=\! [\hat{\mathbf{h}}_{1}(\mathbf{f}_{t,1}), \hat{\mathbf{h}}_{2}(\mathbf{f}_{t,2}), \ldots, \hat{\mathbf{h}}_{N}(\mathbf{f}_{t,N})],
	\end{align}
	where $\hat{\mathbf{h}}_{n}(\mathbf{f}_{t,n}) \!=\! [h_{1,n}(\mathbf{f}_{r,1}),  \ldots, h_{M,n}(\mathbf{f}_{r,M})]^\mathrm{T}$.
	Let $\mathbf{s} \in \mathbb{C}^N$ denote the transmit signal vector. The corresponding covariance matrix is defined as $\mathbf{Q}= \mathbb{E}\{\mathbf{s}\mathbf{s}^H\} \in \mathbb{C}^{N \times N}$, with $\mathbf{Q} \succeq \mathbf{0}$. We consider an average sum power constraint at the transmitter given by $\mathrm{Tr}(\mathbf{Q}) \le P_{\mathrm{max}}$. The received signal vector is thus given by
	\begin{align}
		\mathbf{y}({\mathbf{f}_{t},\mathbf{f}_{r}}) = \mathbf{H}({\mathbf{f}_{t},\mathbf{f}_{r}})\mathbf{s}+\mathbf{n},
	\end{align}
	where $\mathbf{n} \sim \mathcal{CN}(0, \sigma^2 \mathbf{I}_M)$ denotes the additive white Gaussian noise (AWGN) vector at the receiver.
	To investigate the theoretical capacity limit of the RA-enabled MIMO communication system, the perfect channel state information (CSI) is assumed to be available \cite{XueRACSI}. Thus, the MIMO channel capacity can be written as
	\begin{align}
		\label{MIMO_Capacity}
		\!\!\!\! C \!\!= \!\!\!\!\!\! \max_{\substack{\mathbf{Q}: \mathbf{Q} \succeq 0, \\ \mathrm{Tr}(\mathbf{Q}) \leq P_{\mathrm{max}}}} \!\!\!\! \log_2 \! \det \left(\! \mathbf{I}_M \!\!+\!\! \frac{1}{\sigma^2} \mathbf{H}(\mathbf{f}_{t},\mathbf{f}_{r}) \mathbf{Q} \mathbf{H}(\mathbf{f}_{t},\mathbf{f}_{r})^H \!\right).  
	\end{align}
	\subsection{Problem Formulation}
	In this paper, we investigate a MIMO channel capacity maximization problem. The  optimization problem is given by
	\begin{subequations}
		\label{P0}
		\begin{eqnarray}
			\label{P0-0}
			&\!\!\!\!\!\!\!\!\!\!\!\!\!\! \max  \limits_{\mathbf{f}_{t},\mathbf{f}_{r}, \mathbf{Q}}  
			&\!\!\!\! \log_2 \det \left( \mathbf{I}_M + \frac{1}{\sigma^2} \mathbf{H}(\mathbf{f}_{t},\mathbf{f}_{r}) \mathbf{Q} \mathbf{H}(\mathbf{f}_{t},\mathbf{f}_{r})^H \right) \\
			\label{P0-1}
			&\!\!\!\!\!\!\!\!\!\!\!\!\!\!\!\! \mathrm{s.t.}  &\!\!\!\! \mathrm{Tr}(\mathbf{Q}) \leq P_{\mathrm{max}},  \mathbf{Q} \succeq 0, \\
			\label{P0-2}
			&&\!\!\!\! \cos(\theta_{\max}) \le \mathbf{f}_{t,n}^T\mathbf{e}_z \le 1,  \forall n \in \mathcal{N},\\
			\label{P0-3}
			&&\!\!\!\! \cos(\theta_{\max}) \le \mathbf{f}_{r,m}^T\mathbf{e}_z \le 1, \forall m \in \mathcal{M},\\
			\label{P0-4}
			&&\!\!\!\! \Vert \mathbf{f}_{t,n} \Vert_2 = 1, \Vert \mathbf{f}_{r,m} \Vert_2 = 1, \forall n \in \mathcal{N}, \forall m \in \mathcal{M}.
		\end{eqnarray}
	\end{subequations}
	Problem $(\mathrm{\ref{P0}})$ is difficult to solve because the objective function is highly non-concave with respect to the RA orientations. Besides, the feasible set induced by the zenith-angle constraints $(\mathrm{\ref{P0-2}})$, $(\mathrm{\ref{P0-3}})$, and unit-norm $(\mathrm{\ref{P0-4}})$ is non-convex. The coupling between $\mathbf{Q}$, $\mathbf{f}_{t}$, and $\mathbf{f}_{r}$ makes problem $(\mathrm{\ref{P0}})$ more intractable. Therefore, we apply an AO framework to address problem $(\mathrm{\ref{P0}})$ in the following.
	
	\section{Proposed Algorithm}
	\label{Proposed Algorithm}
	In this section, we present an AO algorithm to solve problem $(\mathrm{\ref{P0}})$. 
	Specifically, we first transform the objective function of problem $(\mathrm{\ref{P0}})$ into a more tractable form in terms of the optimization variables in $\mathbf{Q}, \mathbf{f}_{t}$, and $\mathbf{f}_{r}$.
	Then, three subproblems are solved in the sequel, which respectively optimize the transmit covariance matrix $\mathbf{Q}$, the transmit RA orientation $\mathbf{f}_{t}$, and the receive RA orientation $\mathbf{f}_{r}$, with all the other variables being fixed. 
	
	\subsection{Optimization of $\mathbf{Q}$ with given $\mathbf{f}_{t}$ and $\mathbf{f}_{r}$}
	In this subproblem, our objective is to optimize the transmit covariance matrix $\mathbf{Q}$ given the transmit RAs' orientation $\mathbf{f}_{t}$ and the receive RAs' orientation $\mathbf{f}_{r}$. Fixed $\mathbf{H}(\mathbf{f}_{t},\mathbf{f}_{r})$, the optimal solution can be obtained through eigenmode transmission. Specifically, 
	we define $\mathbf{H}(\mathbf{f}_{t},\mathbf{f}_{r}) = \mathbf{U}\mathbf{\Lambda}\mathbf{V}^H$ as the truncated singular value decomposition (SVD) of $\mathbf{H}(\mathbf{f}_{t},\mathbf{f}_{r})$, 
	where $\mathbf{U} \in \mathbb{C}^{M \times S}$, $\mathbf{V} \in \mathbb{C}^{N \times S}$, and $\mathbf{\Lambda} \in \mathbb{C}^{S \times S}$.
	$S=\mathrm{rank}(\mathbf{H}(\mathbf{f}_{t},\mathbf{f}_{r})) \le \min(N,M)$. Therefore, 
	the optimal solution for $\mathbf{Q}$ is given by
	\begin{align}
		\label{Q_optimal}
		\mathbf{Q}^{\star} = {\mathbf{V}} \mathrm{diag}([p_1^{\star}, p_2^{\star}, \dots, p_S^{\star}]) {\mathbf{V}}^H,
	\end{align}
	where the power of the $s$-th data stream, i.e., $p_s^{\star}$, is obtained based on the water-filling principle
	\begin{align}
		\label{Water-filling}
		p_s^{\star} = \max \left( 0, \mu - \frac{\sigma^2}{\boldsymbol{\Lambda}_{s,s}^2} \right), \quad \sum_{s=1}^S p_s^{\star} = P_{\mathrm{max}}.
	\end{align}
	Thus, given $\mathbf{f}_{t}$ and $\mathbf{f}_{r}$, the MIMO channel capacity can be expressed as 
	\begin{align}
		\label{MIMO-optimal}
		C(\mathbf{f}_{t},\mathbf{f}_{r}) = \sum_{s=1}^S \log_2 \left( 1 + \frac{[{\boldsymbol{\Lambda}}]_{s,s}^2 p_s^{\star}}{\sigma^2} \right).
	\end{align}
	
	\subsection{Optimization of $\mathbf{f}_{r}$ with given $\mathbf{Q}$ and $\mathbf{f}_{t}$}
	Denote the SVD of $\mathbf{Q}$ by $\mathbf{Q} = \mathbf{U}_Q\mathbf{\Sigma}_Q\mathbf{U}_Q^H$, with $\mathbf{U}_Q \in \mathbb{C}^{N \times N}$ and $\mathbf{\Sigma}_Q \in \mathbb{C}^{N \times N}$. Since the matrix $\mathbf{Q}$ is positive semidefinite, all diagonal elements of $\mathbf{\Sigma}_Q$ are real numbers that are non-negative. Given $\mathbf{Q}$ and $\{\mathbf{f}_{t,n}\}_{n=1}^N$, we define $\widetilde{\mathbf{H}}(\mathbf{f}_{r}) =\mathbf{H} \mathbf{U}_Q\mathbf{\Sigma}_Q^{\frac{1}{2}} \in \mathbb{C}^{M \times N}$. The $m$-th row of $\widetilde{\mathbf H}(\mathbf f_r)$ corresponds to
	the contribution of the $m$-th receive RA. Then,
	we define its Hermitian transpose as $\widetilde{\mathbf{h}}(\mathbf{f}_{r,m})
	= \widetilde{\mathbf H}(m,:)^H \in\mathbb C^N$. Then, we have
	\begin{align}
		\label{r_m}
		\widetilde{\mathbf{h}}(\mathbf{f}_{r,m}) = \mathbf{\Sigma}_Q^{\frac{1}{2}}\mathbf{U}_Q^H \mathbf{r}(\mathbf{f}_{r,m}),
	\end{align}
	where
	\begin{align}
		\label{r_m0}
		&\mathbf{r}(\mathbf{f}_{r,m}) =[\bar{f}_{m,1}, \bar{f}_{m,2}, \ldots, \bar{f}_{m,N}]^T \in \mathbb{C}^N,\\
		&\bar{a}_{m,n}=a_{m,n}{{f}}_{t,m,n}, \bar{b}_{m,n,d}=b_{m,n,d}{{f}}_{t,n,d},
		\\
		&\bar{f}_{m,n}=\bar{a}_{m,n}{{f}}_{r,m,n} + \sum \nolimits_{d=1}^D \bar{b}_{m,n,d}{{f}}_{r,d,m}.
	\end{align}
	
	Thus, the objective function of problem $(\mathrm{\ref{P0}})$ is written with respect to $\mathbf{f}_{r}$
	\begin{align}
		\label{MIMO2}
		& \log _2 \operatorname{det}\left(\mathbf{I}_M+\frac{1}{\sigma^2} \mathbf{H}(\mathbf{f}_{t},\mathbf{f}_{r}) \mathbf{Q} \mathbf{H}(\mathbf{f}_{t},\mathbf{f}_{r})^H\right) \nonumber  \\
		& =\log _2 \operatorname{det}\left(\mathbf{I}_M+\frac{1}{\sigma^2} \widetilde{\mathbf{H}}(\mathbf{f}_{r}) \widetilde{\mathbf{H}}(\mathbf{f}_{r})^H\right) \nonumber \\
		& \stackrel{(a)}{=} \log _2 \operatorname{det}\left(\mathbf{I}_N+\frac{1}{\sigma^2} \widetilde{\mathbf{H}}(\mathbf{f}_{r})^H \widetilde{\mathbf{H}}(\mathbf{f}_{r})\right) \nonumber \\
		& =\log _2 \operatorname{det}\left(\mathbf{I}_N+\frac{1}{\sigma^2} \sum_{m=1}^M \widetilde{\mathbf{h}}(\mathbf{f}_{r,m}) \widetilde{\mathbf{h}}(\mathbf{f}_{r,m})^H \right),
	\end{align}
	where $(a)$ holds due to $\mathrm{det}(\mathbf{I}_p + \mathbf{A}\mathbf{B}) = \mathrm{det}(\mathbf{I}_q + \mathbf{B}\mathbf{A})$ for $\mathbf{A} \in \mathbb{C}^{p \times q}$ and $\mathbf{B} \in \mathbb{C}^{q \times p}$. 
	Since $\mathbf{H}(\mathbf{f}_{t},\mathbf{f}_{r}) \mathbf{Q} \mathbf{H}(\mathbf{f}_{t},\mathbf{f}_{r})^H$ can be expressed as a summation of $M$ rank-one matrices, the $M$ matrices should be optimally balanced by designing $ \{\mathbf{f}_{r,i}\}_{i=1}^M$ to maximize the MIMO channel capacity.
	
	Although the receive RA orientations are still coupled through the
	determinant in $(\ref{MIMO2})$, this expression enables a block-wise update of
	each receive RA orientation by fixing the others.
	Thus, we can optimize each $\mathbf{f}_{r,m}$ while other variables keep fixed.
	Define $\widetilde{\mathbf{H}}_m \!\!=\!\! [\widetilde{\mathbf{h}}(\mathbf{f}_{r,1}), \ldots, \widetilde{\mathbf{h}}(\mathbf{f}_{r,m-1}),\widetilde{\mathbf{h}}(\mathbf{f}_{r,m+1}), \dots, \widetilde{\mathbf{h}}(\mathbf{f}_{r,M})]^H $. Then, the objective function of problem $(\mathrm{\ref{P0}})$ is written as
	\begin{align}
		& \log _2 \operatorname{det}\left(\mathbf{I}_N+\frac{1}{\sigma^2}\left(\widetilde{\mathbf{H}}_m^H \widetilde{\mathbf{H}}_m+\widetilde{\mathbf{h}}(\mathbf{f}_{r,m}) \widetilde{\mathbf{h}}(\mathbf{f}_{r,m})^H\right)\right) \nonumber  \\
		& \stackrel{\left(b_1\right)}{=}\! \log _2 \!\operatorname{det}\left(\! \!\mathbf{I}_N \!\!+\!\! \frac{1}{\sigma^2}\! \left(\!\mathbf{I}_N \!\!+\!\! \frac{1}{\sigma^2} \widetilde{\mathbf{H}}_m^H \widetilde{\mathbf{H}}_m \! \right)^{-1} \!\!\!\! \widetilde{\mathbf{h}}(\mathbf{f}_{r,m}) \widetilde{\mathbf{h}}(\mathbf{f}_{r,m})^H \!\! \!\right) \nonumber \\
		& \quad +\! \log _2 \operatorname{det}\left(\mathbf{I}_N+\frac{1}{\sigma^2} \widetilde{\mathbf{H}}_m^H \widetilde{\mathbf{H}}_m\right) \nonumber \\
		& \stackrel{\left(b_2\right)}{=} \log _2 \! \left(\! 1\!+\! \frac{1}{\sigma^2} \widetilde{\mathbf{h}}(\mathbf{f}_{r,m})^H \!\!\left(\mathbf{I}_N \!\!+\!\! \frac{1}{\sigma^2} \widetilde{\mathbf{H}}_m^H \widetilde{\mathbf{H}}_m \! \right)^{-1} \!\! \widetilde{\mathbf{h}}(\mathbf{f}_{r,m})\right) \nonumber \\
		& \quad +\log _2 \operatorname{det}\left(\mathbf{I}_N+\frac{1}{\sigma^2} \widetilde{\mathbf{H}}_m^H \widetilde{\mathbf{H}}_m\right), 
	\end{align}
	where $(b_1)$ holds because $\mathrm{det}(\mathbf{A}\mathbf{B}) = \mathrm{det}(\mathbf{A}) \mathrm{det}(\mathbf{B})$ holds for two square matrices $\mathbf{A}$ and $\mathbf{B}$ of equal size. $(b_2)$ holds due to $\mathrm{det}(\mathbf{I}_p + \mathbf{A}\mathbf{B}) = \mathrm{det}(\mathbf{I}_q + \mathbf{B}\mathbf{A})$ for $\mathbf{A} \in \mathbb{C}^{p \times q}$ and $\mathbf{B} \in \mathbb{C}^{q \times p}$. 
	When fixed $\mathbf{Q}$, $ \{\mathbf{f}_{r}\}_{i=1, i \neq m}^M$, and $\{\mathbf{f}_{t,n}\}_{n=1}^N$, the MIMO capacity maximization is equivalent to maximize 
	\begin{align}
		\label{g(r_m)}
		\mathbf{g}(\mathbf{f}_{r,m}) = \widetilde{\mathbf{h}}(\mathbf{f}_{r,m})^H\mathbf{P}_m \widetilde{\mathbf{h}}(\mathbf{f}_{r,m}),
	\end{align}
	
	where $\mathbf{P}_m = \left(\mathbf{I}_N+\frac{1}{\sigma^2} \widetilde{\mathbf{H}}_m^H \widetilde{\mathbf{H}}_m\right)^{-1} \in \mathbb{C}^{N \times N}$ is a positive definite matrix independent of $\mathbf{f}_{r,m}$. To reduce the computational complexity of matrix inverse operation for $1 \le m \le M$, matrix ${\mathbf{P}}_m$ can be updated based on ${\mathbf{P}}_{m-1}$. Specifically, we define $\mathbf{Z}_1 = [\widetilde{\mathbf{h}}(\mathbf{f}_{r,m-1}), \widetilde{\mathbf{h}}(\mathbf{f}_{r,m})] \in \mathbb{C}^{N \times 2}$ and $\mathbf{Z}_2 = [\widetilde{\mathbf{h}}(\mathbf{f}_{r,m-1}), -\widetilde{\mathbf{h}}(\mathbf{f}_{r,m})] \in \mathbb{C}^{N \times 2}$ such that
	\begin{equation}
		\widetilde{\mathbf{H}}_{m}^H \widetilde{\mathbf{H}}_{m} = \widetilde{\mathbf{H}}_{m-1}^H \widetilde{\mathbf{H}}_{m-1} + \mathbf{Z}_1 \mathbf{Z}_2^H. 
	\end{equation}
	Since $(\mathbf{A}+\mathbf{BC})^{-1} = \mathbf{A}^{-1}-\mathbf{A}^{-1}\mathbf{B}(\mathbf{I}_q+\mathbf{CA}^{-1}\mathbf{B})^{-1}\mathbf{CA}^{-1}$  for $\mathbf{A} \in \mathbb{C}^{p \times p}, \mathbf{B} \in \mathbb{C}^{p \times q}$ and $\mathbf{C} \in \mathbb{C}^{q \times p}$, we have
	\begin{align}
		\label{P_m}
		&\!\!\!\! \mathbf{P}_m = \mathbf{P}_{m-1} \nonumber \\
		&\!\!\!\!-\frac{1}{\sigma^2} \mathbf{P}_{m-1}
		\mathbf{Z}_1 \left( \mathbf{I}_2 + \frac{1}{\sigma^2} \mathbf{Z}_2^H \mathbf{P}_{m-1} \mathbf{Z}_1 \right)^{-1} \mathbf{Z}_2^H \mathbf{P}_{m-1}.
	\end{align}
	Furthermore, we define a positive semidefinite matrix
	\begin{equation}
		\mathbf{B}_m \triangleq \mathbf{U}_Q \mathbf{\Sigma}_Q^{\frac{1}{2}} \mathbf{P}_m \mathbf{\Sigma}_Q^{\frac{1}{2}} \mathbf{U}_Q^H  \in \mathbb{C}^{N \times N}.
	\end{equation}
	According to (\ref{r_m0}), $\mathbf{g}(\mathbf{f}_{r,m})$ in (\ref{g(r_m)}) can be rewritten in the following form with respect to $\mathbf{h}(\mathbf{f}_{r,m})$
	\begin{equation}
		\begin{aligned}
			\mathbf{g}(\mathbf{f}_{r,m}) & = \mathbf{r}(\mathbf{f}_{r,m})^H \mathbf{U}_Q \mathbf{\Sigma}_Q^{\frac{1}{2}} \mathbf{P}_m \mathbf{\Sigma}_Q^{\frac{1}{2}} \mathbf{U}_Q^H \mathbf{r}(\mathbf{f}_{r,m}) \\
			& = \mathbf{r}(\mathbf{f}_{r,m})^H \mathbf{B}_m \mathbf{r}(\mathbf{f}_{r,m}).
		\end{aligned}
	\end{equation}
	Since $\mathbf{B}_m$ is a constant matrix independent of $\mathbf{f}_{r,m}$, the subproblem for optimizing $\mathbf{f}_{r,m}$ is given by
	\begin{subequations}
		\label{P1}
		\begin{eqnarray}
			\label{P1-0}
			&\!\!\!\!\!\!\! \max  \limits_{\mathbf{f}_{r,m}}  
			& \mathbf{r}(\mathbf{f}_{r,m})^H \mathbf{B}_m \mathbf{r}(\mathbf{f}_{r,m}) \\
			\label{P1-1}
			&\!\!\!\!\!\!\! \mathrm{s.t.}  & \Vert \mathbf{f}_{r,m} \Vert_2 = 1, \\
			\label{P1-2}
			&& \cos(\theta_{\max}) \le \mathbf{f}_{r,m}^T\mathbf{e}_z \le 1.
		\end{eqnarray}
	\end{subequations}
	The objective function of problem $(\mathrm{\ref{P1}})$ is a non-concave function for $\mathbf{f}_{r,m}$. Moreover, constraints $(\mathrm{\ref{P1-1}})$ and $(\mathrm{\ref{P1-2}})$ are non-convex, which makes problem $(\mathrm{\ref{P1}})$ still non-convex.
	
	To address problem \eqref{P1}, we first define the feasible set as the spherical cap
	\begin{equation}
		\label{eq:spherical-cap}
		\mathcal{C} = \{ \mathbf{x}\in \mathbb{R}^3 : \Vert \mathbf{x}\Vert_2 = 1, \ \mathbf{x}^T \mathbf{e}_z \geq \cos(\theta_{\max}) \}.
	\end{equation} 
	Then, we apply the Riemannian Frank-Wolfe algorithm for solving problem $(\mathrm{\ref{P1}})$. Let $F(\mathbf{f}_{r,m}) = \mathbf{r}(\mathbf{f}_{r,m})^H \mathbf{B}_m \mathbf{r}(\mathbf{f}_{r,m})$, $\mathbf{s}_{d,m} = \mathbf{s}_d-{\mathbf{r}}_{m}$, and $\mathbf{d}_{m,n} = \mathbf{t}_n-{\mathbf{r}}_{m}$. The $n$-th component of $\mathbf{r}(\mathbf{f}_{r,m})$ is given by
	\begin{align}
		\!\!\!\!  \bar{f}_{m,n} = \bar{a}_{m,n} f_{r,m,n} + \sum\nolimits_{d=1}^{D} \bar{b}_{m,n,d} f_{r,d,m}.
	\end{align}
	We first choose an initial feasible point $\mathbf{f}_{r,m}^{(t)} \in \mathcal{C}$,
	The Euclidean gradient with respect to $\mathbf{f}_{r,m}$ is given by
	\begin{align}
		\frac{\partial F(\mathbf{f}_{r,m}^{(t)})}{\partial \mathbf{f}_{r,m}^{(t)}}  = 2 \Re\left( \mathbf{J}_{\mathbf{f}_{r,m}^{(t)}}^H \mathbf{B}_m \mathbf{r}(\mathbf{f}_{r,m}^{(t)}) \right), \label{eq:euclidean-gradient}
	\end{align}
	where $\mathbf{J}_{\mathbf{f}_{r,m}^{(t)}} \in \mathbb{C}^{N \times 3}$ is the Jacobian matrix of $\mathbf{r}(\mathbf{f}_{r,m})$ with respect to $\mathbf{f}_{r,m}$. The $n$-th row of the Jacobian matrix is
	
	\begin{equation}
		\label{eq:jacobian_row}
		\frac{\partial [\mathbf{r}]_n}{\partial \mathbf{f}_{r,m}^{(t)}} = \bar{a}_{m,n} \frac{\partial f_{r,m,n}}{\partial \mathbf{f}_{r,m}^{(t)}} + \sum_{d=1}^{D} \bar{b}_{m,n,d} \frac{\partial f_{r,d,m}}{\partial \mathbf{f}_{r,m}^{(t)}},
	\end{equation}
	where
	\begin{align}
		\!\!\frac{\partial f_{r,m,n}}{\partial \mathbf{f}_{r,m}^{(t)}} &\!\!= \!\!
		\begin{cases}
			p \left(\! \dfrac{\widetilde{\mathbf{f}}_{r,m}^{(t),T} \mathbf{d}_{m,n}}{r_{m,n}} \!\right)^{p-1}  \dfrac{\mathbf{R}_r^T\mathbf{d}_{m,n}}{r_{m,n}}, \!\!&\!\! \text{if } \widetilde{\mathbf{f}}_{r,m}^{(t),T} \mathbf{d}_{m,n} \!\! >\!\! 0, \\
			\bm{0}_{3 \times 1}, & \text{otherwise},
		\end{cases} \label{eq:grad_direct} 
	\end{align}
	\begin{align}
		\!\!\frac{\partial f_{r,d,m}}{\partial \mathbf{f}_{r,m}^{(t)}} &\!\!= \!\!
		\begin{cases}
			p \left(\! \dfrac{\widetilde{\mathbf{f}}_{r,m}^{(t),T} \mathbf{s}_{d,m}}{r_{d,m}} \!\right)^{p-1}  \dfrac{\mathbf{R}_r^T\mathbf{s}_{d,m}}{r_{d,m}}, \!\!&\!\! \text{if } \widetilde{\mathbf{f}}_{r,m}^{(t),T} \mathbf{s}_{d,m} \!\!>\!\! 0, \\
			\bm{0}_{3 \times 1}, & \text{otherwise},
		\end{cases} \label{eq:grad_scatter}
	\end{align}
	where $\widetilde{\mathbf{f}}_{r,m}^{(t)} = \mathbf{R}_r\mathbf{f}_{r,m}^{(t)}$.
	Then, for the unit-norm constraint \eqref{P1-1}, we project $\frac{\partial F(\mathbf{f}_{r,m}^{(t)})}{\partial \mathbf{f}_{r,m}^{(t)}}$ into the tangent space of the unit sphere $\mathbf{f}_{r,m}^{(t)}$
	\begin{align}
		\mathbf{g}_{r,m}^{(t)} = \left( \mathbf{I} - \mathbf{f}_{r,m}^{(t)}(\mathbf{f}_{r,m}^{(t)})^T \right) \frac{\partial F(\mathbf{f}_{r,m}^{(t)})}{\partial \mathbf{f}_{r,m}^{(t)}}. \label{eq:riemannian-gradient}
	\end{align}
	Then, we select an orientation on the spherical cap by maximizing the linearized utility along $\mathbf{g}_{r,m}^{(t)} $
	\begin{align}
		\mathbf{y}_{r,m}^{(t)} = \mathrm{argmax}_{\mathbf{x}_{r,m} \in \mathcal{C}}  \mathbf{g}_{r,m}^{H,(t)} \mathbf{x}_{r,m}. \label{eq:fw-subproblem}
	\end{align}
	Define
	\begin{align}
		\hat{\mathbf{g}}_{r,m} &= \frac{\mathbf{g}_{r,m}^{(t)}}{\Vert {\mathbf{g}_{r,m}^{(t)}}\Vert_2},  c_z = \cos(\theta_{\max}),  z_{r,m}^{(t)} = \hat{\mathbf{g}}_{r,m}^T \mathbf{e}_z.
	\end{align}
	The closed-form solution of $\mathbf{y}_{r,m}^{(t)}$ can be obtained as 
	\begin{align}
		\!\!	\mathbf{y}_{r,m}^{(t)} \!\!= \!\!
		\begin{cases}
			\hat{\mathbf{g}}_{r,m}^{(t)}, \!\! & \text{if } z_{r,m}^{(t)} \geq c_z, \\
			\sqrt{1 \!-\! c_z^2} \, \mathbf{v}_{r,m}^{(t)} \!+\! c_z \mathbf{e}_z, \!\! & \text{if } z_{r,m}^{(t)} \!\!<\!\! c_z \text{ and } \Vert \mathbf{v}_{r,m}^{(t)} \Vert_2 \!>\! 0, \\[6pt]
			\sqrt{1 \!-\! c_z^2} \, \mathbf{u} \!+\! c_z \mathbf{e}_z, \!\! & \text{if } z_{r,m}^{(t)} \!\!<\!\! c_z \text{ and } \Vert \mathbf{v}_{r,m}^{(t)} \Vert_2 \!=\! 0,
		\end{cases} \label{eq:closed-form-y}
	\end{align}
	where $\mathbf{v}_{r,m}^{(t)} = (\hat{\mathbf{g}}_{r,m}^{(t)} - z_{r,m}^{(t)}\mathbf{e}_z)/\Vert \hat{\mathbf{g}}_{r,m}^{(t)} - z_{r,m}^{(t)}\mathbf{e}_z \Vert_2$, and $\mathbf{u}$ is any unit vector orthogonal to $\mathbf{e}_z$.
	The search direction of $\mathbf{f}_{r,m}$ is $\mathbf{d}_{r,m}^{(t)} = \mathbf{y}_{r,m}^{(t)} - \mathbf{f}_{r,m}^{(t)}$.
	
	Initialize step size $\rho = 1$, reduction factor $\beta \in (0,1)$, and constant $c \in (0,1)$. While the Armijo condition is not satisfied
	\begin{align}
		& \!\!\!\!\! F\left( \frac{\mathbf{f}_{r,m}^{(t)} \!+\! \rho^{(t)} \mathbf{d}_{r,m}^{(t)}}{\Vert \mathbf{f}_{r,m}^{(t)} \!+\! \rho^{(t)} \mathbf{d}_{r,m}^{(t)} \Vert_2} \right) \!\geq \! F(\mathbf{f}_{r,m}^{(t)}) \!+ \! c \rho {\mathbf{g}}_{r,m}^{H,(t)}\mathbf{d}_{r,m}^{(t)}, \label{eq:armijo}
	\end{align}
	set $\rho \leftarrow \beta \rho$. Then set $\rho^{(t)} = \rho$.
	The orientation can be updated by a retraction
	\begin{equation}
		\mathbf{f}_{r,m}^{(t+1)} = \frac{\mathbf{f}_{r,m}^{(t)} + \rho^{(t)} \mathbf{d}_{r,m}^{(t)}}{\Vert \mathbf{f}_{r,m}^{(t)} + \rho^{(t)} \mathbf{d}_{r,m}^{(t)} \Vert_2}, \label{eq:retraction}
	\end{equation}
	where $ \rho^{(t)} \in (0,1]$ is a step size that can be selected by a standard backtracking line search to guarantee a sufficient increase of the utility $F(\mathbf{f}_{r,m})$. The detailed procedure is presented in Algorithm \ref{alg:fw-ra}.
	
	Then, we consider the special case with \(p=1\). Specifically, we first define
	\begin{equation}
		\hat{\mathbf{a}}_{m,n}
		\triangleq
		\bar a_{m,n}\frac{\mathbf t_n-\mathbf r_m}{r_{m,n}}
		+
		\sum_{d=1}^{D}
		\bar b_{m,n,d}
		\frac{\mathbf s_d-\mathbf r_m}{r_{d,m}}
		\in \mathbb C^{3\times 1}.
	\end{equation}
	Then, the \(n\)-th element of \(\mathbf r(\mathbf f_{r,m})\) can be
	rewritten as
	\begin{equation}
		[\mathbf r(\mathbf f_{r,m})]_n
		=
		(\mathbf R_r\mathbf f_{r,m})^T
		\hat{\mathbf{a}}_{m,n}.
	\end{equation}
	Let $\mathbf A_m
	=
	[\hat{\mathbf{a}}_{m,1},\hat{\mathbf{a}}_{m,2},\ldots,
	\hat{\mathbf{a}}_{m,N}]
	\in \mathbb C^{3\times N}$.
	We have
	\begin{equation}
		\label{r-equivalent}
		\mathbf r(\mathbf f_{r,m})
		=
		\mathbf A_m^T\mathbf R_r\mathbf f_{r,m}.
	\end{equation}
	Substituting \eqref{r-equivalent} into (35a), we have
	\begin{align}
		\mathbf r^H(\mathbf f_{r,m})\mathbf B_m\mathbf r(\mathbf f_{r,m})
		&=
		(\mathbf R_r\mathbf f_{r,m})^H
		\mathbf A_m^*
		\mathbf B_m
		\mathbf A_m^T
		(\mathbf R_r\mathbf f_{r,m})
		\nonumber\\
		&=
		(\mathbf R_r\mathbf f_{r,m})^T
		\mathbf C_m
		(\mathbf R_r\mathbf f_{r,m}),
	\end{align}
	where
	\begin{equation}
		\mathbf C_m
		\triangleq
		\operatorname{Re}\left\{
		\mathbf A_m^*
		\mathbf B_m
		\mathbf A_m^T
		\right\}
		\in \mathbb R^{3\times 3}.
	\end{equation}
	Therefore, problem (35) reduces to
	\begin{subequations}
		\label{P1-3}
		\begin{eqnarray}
			\label{P1-3-0}
			&\!\!\!\!\!\!\! \max  \limits_{\mathbf{f}_{r,m}}  
			&\!\!\!\! (\mathbf{R}_r\mathbf{f}_{r,m})^\mathrm{T}\mathbf C_m(\mathbf{R}_r\mathbf{f}_{r,m}) \\
			\label{P1-3-1}
			&\!\!\!\!\!\!\! \mathrm{s.t.} &\!\!\!\! \eqref{P1-1}, \eqref{P1-2}.
		\end{eqnarray}
	\end{subequations}
	When omitting constraint $(\mathrm{\ref{P1-2}})$, the best $\mathbf{R}_r\mathbf{f}_{r,m}$ is the maximum eigenvector of $\mathbf C_m$, i.e., $\mathbf{a}_0$. Thus, $\mathbf{f}_{r,m} = \mathbf{R}_r^T\mathbf{a}_0=[a_x,a_y,a_z]$. Then, we have
	\begin{align}
		\theta_{z,m}^{r} = \arccos\left({a_z}/{d_m}\right),
	\end{align}
	where $d_m = \sqrt{a_x^2+a_y^2+a_z^2}$.
	When $\theta_{z,m}^{r} \le \theta_{\max}$, the optimal $\mathbf{f}_{r,m}^\mathrm{\star}$ is $ \mathbf{R}_r^T\mathbf{a}_0$. However, if $\theta_{z,m}^{r} > \theta_{\max}$, we can apply Algorithm \ref{alg:fw-ra} to obtain the solution. 
	
	\begin{algorithm}
		\caption{Frank-Wolfe Algorithm for Solving Problem $(\mathrm{\ref{P1}})$}
		\label{alg:fw-ra}
		\begin{algorithmic}[1]
			\Require Initial orientations $\{{\mathbf{f}}_{r,m}^{(0)}\}$, tolerance $\epsilon > 0$, maximum iterations $T_{\max}$;
			\State $t = 0$;
			\Repeat
			\State Compute Riemannian gradient $\mathbf{g}_{r,m}^{(t)}$ via $(\mathrm{\ref{eq:riemannian-gradient}})$;
			\State Solve linear subproblem for $\mathbf{y}_{r,m}^{(t)}$ via $(\mathrm{\ref{eq:fw-subproblem}})$;
			\State Compute search direction $\mathbf{d}_{r,m}^{(t)} = \mathbf{y}_{r,m}^{(t)} - {\mathbf{f}}_{r,m}^{(t)}$;
			\State Find stepsize $\rho^{(t)}$ via Armijo backtracking line search;
			\State Update orientation via \eqref{eq:retraction};
			\State Compute $F^{(t+1)} = F({\mathbf{f}}_{r,m}^{(t+1)})$;
			\State $t = t + 1$;
			\Until{$t \geq T_{\max}$ or $|F^{(t)} - F^{(t-1)}| \leq \epsilon$};
			\State \textbf{Return} Converged $\{\tilde{\mathbf{f}}_{r,m}^{\star}\}$.
		\end{algorithmic}
	\end{algorithm}
	
	\begin{algorithm}[t]
		\caption{AO Algorithm for Solving Problem \eqref{P0}}
		\label{OverallAO}
		\begin{algorithmic}[1]
			\Require Initial feasible transmit RA orientations $\mathbf{f}_t^{(0)}$, 
			initial feasible receive RA orientations $\mathbf{f}_r^{(0)}$, 
			maximum AO iteration number $L_{\max}$, tolerance $\varepsilon$;
			\State Initialize outer iteration index $\ell=0$;
			\State Compute the effective channel $\mathbf{H}(\mathbf{f}_t^{(0)},\mathbf{f}_r^{(0)})$;
			\State Compute the initial objective value $C^{(0)}$ via \eqref{MIMO_Capacity};
			\Repeat
			\State Fixed $\mathbf{f}_t^{(\ell)}$ and $\mathbf{f}_r^{(\ell)}$, update $\mathbf{Q}^{(\ell+1)}$ via \eqref{Q_optimal}-\eqref{MIMO-optimal};
			\For{$m=1$ to $M$}
			\State Fixed $\mathbf{Q}^{(\ell+1)}$, $\mathbf{f}_t^{(\ell)}$, and $\{\mathbf{f}_{r,i}\}_{i\neq m}$, update $\mathbf{f}_{r,m}^{(\ell+1)}$ by  Algorithm~\ref{alg:fw-ra};
			\EndFor
			\State Construct $\mathbf S^{(\ell+1)}$ by \eqref{S-get};
			\For{$n=1$ to $N$}
			\State Fixed $\mathbf S^{(\ell+1)}$, $\mathbf{f}_r^{(\ell+1)}$, and $\{\mathbf{f}_{t,i}\}_{i\neq n}$, update $\mathbf{f}_{t,n}^{(\ell+1)}$ by Algorithm~\ref{alg:fw-ra};
			\EndFor
			\State Compute the updated objective value $C^{(\ell+1)}$ via \eqref{MIMO_Capacity};
			\State $\ell \leftarrow \ell+1$;
			\Until{$\ell\ge L_{\max}$ \textbf{or} $\left|C^{(\ell)}-C^{(\ell-1)}\right|\le \varepsilon$};
			\State \Return Converged $\mathbf{Q}^\star=\mathbf{Q}^{(\ell)}$, $\mathbf{f}_t^\star=\mathbf{f}_t^{(\ell)}$, and $\mathbf{f}_r^\star=\mathbf{f}_r^{(\ell)}$.
		\end{algorithmic}
	\end{algorithm}

	\subsection{Optimization of $\mathbf{f}_{t}$ with given $\mathbf{Q}$ and $ \mathbf{f}_{r}$}
	Since the singular values of channel matrices $\mathbf{H}(\mathbf{f}_{t},\mathbf{f}_{r})$ and $\mathbf{H}(\mathbf{f}_{t},\mathbf{f}_{r})^H$ are the same, the channel capacity of $\mathbf{H}(\mathbf{f}_{t},\mathbf{f}_{r})$ is always equal to that of $\mathbf{H}(\mathbf{f}_{t},\mathbf{f}_{r})^H$ under the same transmit power constraint. Then, we have
	\begin{align}
		&\!\!\!\! \max_{\substack{\mathbf{Q}: \operatorname{Tr}(\mathbf{Q}) \!\leq\! P_{\mathrm{max}}, \\ \mathbf{Q} \succeq \mathbf{0}}} \! \log _2 \operatorname{det}\left(\! \mathbf{I}_M \!+\! \frac{1}{\sigma^2} \mathbf{H}(\mathbf{f}_{t},\mathbf{f}_{r}) \mathbf{Q} \mathbf{H}(\mathbf{f}_{t},\mathbf{f}_{r})^H \!\right) \nonumber \\
		&\!\!\!\! =\!\! \max_{\substack {\mathbf{S}: \operatorname{Tr}(\mathbf{S}) \!\leq\! P_{\mathrm{max}}, \\ \mathbf{S} \succeq \mathbf{0}}} \!\!\!\! \log _2 \!\! \operatorname{det}\left(\! \mathbf{I}_N \!\!+\!\! \frac{1}{\sigma^2} \mathbf{H}(\mathbf{f}_{t},\mathbf{f}_{r})^H \mathbf{S} \mathbf{H}(\mathbf{f}_{t},\mathbf{f}_{r}) \! \right),
	\end{align}
	where $\mathbf{S} \in \mathbb{C}^{M \times M}$ is the equivalent covariance matrix of transmit signals for $\mathbf{H}(\mathbf{f}_{t},\mathbf{f}_{r})^H$, with $\operatorname{Tr}(\mathbf{S}) \leq P_{\mathrm{max}}$ and $ \mathbf{S} \succeq \mathbf{0}$. Similar to the optimization of $\mathbf{Q}$ with given $ \{\mathbf{f}_{r,m}\}_{m=1}^M$ and $\{\mathbf{f}_{t,n}\}_{n=1}^N$, the optimal $\mathbf{S}$ is given by the eigenmode transmission. Let $\mathbf{H}(\mathbf{f}_{t},\mathbf{f}_{r})^H={\mathbf{V}}{\mathbf{\Lambda}}{\mathbf{U}}^H$ as the truncated SVD of $\mathbf{H}(\mathbf{f}_{t},\mathbf{f}_{r})^H$. Then, the optimal $\mathbf{S}$ is given by
	\begin{align}
		\label{S-get}
		\mathbf{S}^{\star} = {\mathbf{U}} \mathrm{diag}([p_1^{\star}, p_2^{\star}, \dots, p_S^{\star}]) {\mathbf{U}}^H,
	\end{align}
	where the power of the $s$-th data stream, i.e., $p_s^{\star}$, is obtained based on the water-filling principle.
	
	Similar to the procedure for optimizing $\mathbf{f}_{r,m}$, we denote $\mathbf{S}=\mathbf{U}_S\mathbf{\Sigma}_S\mathbf{U}_S^H$ as the SVD of $\mathbf{S}$, with $\mathbf{U}_S \in \mathbb{C}^{M \times M}$ and $\mathbf{\Sigma}_S \in \mathbb{C}^{M \times M}$. As the matrix $\mathbf{S}$ is positive semi-definite, all diagonal elements of $\mathbf{\Sigma}_S$ are real numbers that are nonnegative. Then, we define $\widehat{\mathbf{H}}(\mathbf{f}_{t})=\mathbf{H}(\mathbf{f}_{t},\mathbf{f}_{r})^H\mathbf{U}_S\mathbf{\Sigma}_S^{\frac{1}{2}} \in \mathbb{C}^{N \times M}$. The \(n\)-th row of \(\widehat{\mathbf H}(\mathbf f_t)\) corresponds to
	the contribution of the \(n\)-th transmit RA. We define its Hermitian
	transpose as $\widehat{\mathbf h}(\mathbf f_{t,n})=\widehat{\mathbf H}(n,:)^H
	\in\mathbb C^M$. Then, we have $\hat{\mathbf{h}}(\mathbf{f}_{t,n}) =\mathbf{\Sigma}_S^{\frac{1}{2}}\mathbf{U}_S^H\mathbf{t}(\mathbf{f}_{t,n}) \in \mathbb{C}^M$, where 
	\begin{align}
		\label{t_n}
		&\mathbf{t}(\mathbf{f}_{t,n}) =[\tilde{f}_{1,n}, \tilde{f}_{2,n}, \ldots, \tilde{f}_{M,n}]^T \in \mathbb{C}^M,\\
		\label{r_m}
		&\tilde{a}_{m,n}=a_{m,n}{{f}}_{r,m,n}, \tilde{b}_{m,n,d}=b_{m,n,d}{{f}}_{r,d,m},
		\\
		&\tilde{f}_{m,n}=\tilde{a}_{m,n}{{f}}_{t,m,n} + \sum \nolimits_{d=1}^D \tilde{b}_{m,n,d}{{f}}_{t,n,d}.
	\end{align}
	Then, we remove $\mathbf{t}(\mathbf{f}_{t,n})$ from $\widehat{\mathbf{H}}(\mathbf{f}_{t})$ and denote the remaining $M \times (N-1)$ sub-matrix by $\widehat{\mathbf{H}}_n=[\hat{\mathbf{h}}(\mathbf{f}_{t,1}), \ldots, \hat{\mathbf{h}}(\mathbf{f}_{t,n-1}),\hat{\mathbf{h}}(\mathbf{f}_{t,n+1}), \dots, \hat{\mathbf{h}}(\mathbf{f}_{t,N})]^H$. The MIMO capacity maximization is equivalent to maximize 
	\begin{align}
		\hat{\mathbf{g}}(\mathbf{f}_{t,n}) = \hat{\mathbf{h}}(\mathbf{f}_{t,n})^H\mathbf{D}_n \hat{\mathbf{h}}(\mathbf{f}_{t,n}),
	\end{align}
	where $\mathbf{D}_n = \left(\boldsymbol{I}_M+\frac{1}{\sigma^2} \widehat{\mathbf{H}}_n^H \widehat{\mathbf{H}}_n\right)^{-1} \in \mathbb{C}^{M \times M}$, which can be updated similarly to $(\ref{P_m})$. Next, we define
	\begin{equation}
		\widehat{\mathbf{B}}_n \triangleq \mathbf{U}_S \mathbf{\Sigma}_S^{\frac{1}{2}} \mathbf{D}_n \mathbf{\Sigma}_S^{\frac{1}{2}} \mathbf{U}_S^H  \in \mathbb{C}^{M \times M}.
	\end{equation}
	Thus, the optimization of $\mathbf{f}_{t,n}$ is expressed as
	\begin{subequations}
		\label{P2}
		\begin{eqnarray}
			\label{P2-0}
			&\!\!\!\!\!\!\! \max  \limits_{\mathbf{f}_{t,n}}  
			& \mathbf{t}(\mathbf{f}_{t,n})^H \widehat{\mathbf{B}}_n \mathbf{t}(\mathbf{f}_{t,n}) \\
			\label{P2-1}
			&\!\!\!\!\!\!\! \mathrm{s.t.}  & \Vert \mathbf{f}_{t,n} \Vert_2 = 1, \\
			\label{P2-2}
			&& \cos(\theta_{\max}) \le \mathbf{f}_{t,n}^T\mathbf{e}_z \le 1.
		\end{eqnarray}
	\end{subequations}
	Since problem \eqref{P2} has the same structure as problem \eqref{P1}, problem \eqref{P2} can be solved in a similar way using Algorithm \ref{alg:fw-ra}. The details are omitted here.

	\subsection{Computational Complexity and Convergence Analysis}
	The complete procedure for solving problem~\eqref{P0} is summarized in Algorithm~\ref{OverallAO}.
	Note that in each AO iteration, the transmit covariance matrix update gives the optimal solution of $\mathbf Q$ with fixed antenna orientations, while each transmit/receive RA orientation is updated by Algorithm~\ref{alg:fw-ra} to yield a non-decreasing objective value through the Armijo backtracking line search. Thus, the channel capacity is monotonically non-decreasing over the AO iterations. On the other hand, under the transmit power constraint $\mathrm{Tr}(\mathbf Q)\le P_{\mathrm{max}}$, the achievable channel capacity is upper bounded. Therefore, the objective sequence generated by the proposed AO algorithm is guaranteed to converge. Since the original problem is non-convex, the proposed algorithm generally converges to a locally optimal solution of problem~\eqref{P0}.
	
	The computational complexity of the proposed AO algorithm mainly comes from three parts: 1) updating the transmit covariance matrix $\mathbf Q$, 2) updating all receive RA orientations, and 3) updating all transmit RA orientations.
	For fixed $\mathbf f_t$ and $\mathbf f_r$, updating $\mathbf Q$ requires the truncated SVD of the effective channel matrix $\mathbf H(\mathbf f_t,\mathbf f_r)\in\mathbb C^{M\times N}$, whose complexity is $\mathcal{C}_Q=\mathcal O\left(MN\min(M,N)\right)$. 
	For the receive-side update, let $I_r$ denote the average number of inner iterations for updating one receive RA orientation. Since each iteration mainly involves Jacobian evaluation and matrix-vector operations, the total complexity for updating all $M$ receive RAs is $\mathcal{C}_R=\mathcal O\left(MI_r(N^2+ND)\right)$. Similarly, letting $I_t$ denote the average number of inner iterations for updating one transmit RA orientation, the total complexity for updating all $N$ transmit RAs is $\mathcal{C}_T=\mathcal O\left(NI_t(M^2+MD)\right)$.
	Therefore, with $I_{\mathrm{AO}}$ outer AO iterations, the overall complexity of the proposed algorithm is given by $\mathcal{C}_{\mathrm{all}} = \mathcal O\Big(
	I_{\mathrm{AO}}(\mathcal{C}_Q + \mathcal{C}_R + \mathcal{C}_T)\Big)$.

	\subsection{Solution for Problem $(\ref{P0})$ in the low-SNR regime}
	\label{Solution for low-SNR regime}
	In this subsection, we investigate the low-SNR case, which may occur with low transmission power and/or long distance between the transmitter and the receiver. In this regime, the optimal strategy is beamforming over the strongest eigenmode of the effective MIMO channel $\mathbf{H}$ by allocating all transmit power to the strongest eigenchannel. Specifically, the optimal transmit covariance matrix that maximizes the capacity is given by $\mathbf{Q}=P_{\mathrm{max}}\mathbf{u}_Q\mathbf{u}_Q^H$, where $\mathbf{u}_Q \in \mathbb{C}^N$ denotes the transmit beamforming vector. Thus, the capacity in the low-SNR regime can be represented as
	\begin{align}
		\label{C_L}
		&  C_L(\mathbf{f}_{t}, \mathbf{f}_{r})= \nonumber \\
		&  \max_{\mathbf{u}_Q:\left\|\mathbf{u}_Q\right\|_2^2=1} \!\!\! \log_2 \! \operatorname{det}\left(\! \mathbf{I}_M \!\! + \!\! \frac{P_{\mathrm{max}}}{\sigma^2} \mathbf{H}(\mathbf{f}_{t}, \mathbf{f}_{r}) \mathbf{u}_Q\mathbf{u}_Q^H \mathbf{H}(\mathbf{f}_{t}, \mathbf{f}_{r})^H \!\right) \nonumber \\
		& \stackrel{(c)}{=}\!\!\!\!\!\! \max_{\mathbf{u}_Q:\left\|\mathbf{u}_Q\right\|_2^2=1} \log _2\left(1+\frac{P_{\mathrm{max}}}{\sigma^2}\left\|\mathbf{H}(\mathbf{f}_{t}, \mathbf{f}_{r}) \mathbf{u}_Q\right\|_2^2\right),
	\end{align}
	where $(c)$ holds due to $\mathrm{det}(\mathbf{I}_p + \mathbf{A}\mathbf{B}) = \mathrm{det}(\mathbf{I}_q + \mathbf{B}\mathbf{A})$ for $\mathbf{A} \in \mathbb{C}^{p \times q}$ and $\mathbf{B} \in \mathbb{C}^{q \times p}$. Based on $(\ref{C_L})$, maximizing channel capacity is equivalent to maximizing $\Vert \mathbf{H}(\mathbf{f}_{t}, \mathbf{f}_{r}) \mathbf{u}_Q \Vert_2^2$. Fixed $\mathbf{H}(\mathbf{f}_{t}, \mathbf{f}_{r})$, the optimal $\mathbf{u}_Q$ is the strongest right singular vector of $\mathbf{H}(\mathbf{f}_{t}, \mathbf{f}_{r})$. $\Vert \mathbf{H}(\mathbf{f}_{t}, \mathbf{f}_{r}) \mathbf{u}_Q \Vert_2^2 = \sum_{m=1}^{M}\vert \mathbf{u}_Q^H \mathbf{r}_m(\mathbf{f}_{r,m}) \vert^2$. Therefore, given $\mathbf{u}_Q$, $ \{\mathbf{f}_{r,i}\}_{i=1, i \neq m}^M$ and $\{\mathbf{f}_{t,n}\}_{n=1}^N$, maximizing the channel capacity is equivalent to maximize $\vert \mathbf{u}_Q^H \mathbf{r}_m(\mathbf{f}_{r,m}) \vert^2$ for each $\mathbf{f}_{r,m}$. The subproblem for optimizing $\mathbf{f}_{r,m}$ can be formulated similarly to problem $(\ref{P1})$ by replacing $\mathbf{B}_m$ by $\mathbf{u}_Q\mathbf{u}_Q^H$. Similarly, fixed $\mathbf{u}_Q$, $ \{\mathbf{f}_{r,m}\}_{m=1}^M$ and $\{\mathbf{f}_{t,i}\}_{i=1, i \neq n}^N$, the optimal $\mathbf{S}$ is given by $\mathbf{S}=P_{\mathrm{max}}\mathbf{u}_S\mathbf{u}_S^H$, where $\mathbf{u}_S \in \mathbb{C}^M$ is the strongest right singular vector of $\mathbf{H}(\mathbf{f}_{t}, \mathbf{f}_{r})^H$. The subproblem for optimizing $\mathbf{f}_{t,n}$ is expressed similarly to problem $(\ref{P2})$ by replacing ${\mathbf{D}}_n$ by $\mathbf{u}_S\mathbf{u}_S^H$. Thus, by iteratively optimizing one variable with other variables fixed, Algorithm \ref{alg:fw-ra} can be applied to obtain a locally optimal solution for maximizing channel capacity in the low-SNR regime.
	
	\subsection{Solution to Problem $(\ref{P0})$ With Single-Antenna Transmitter/Receiver}
	We have investigated problem $(\ref{P0})$ for the general MIMO channel with $M \ge 1$ and $N \ge 1$, by considering parallel transmissions of multiple data streams in general. In this subsection, we consider problem $(\ref{P0})$ for the special cases with $N =1$ or $M=1$, where only one data stream can be transmitted. This leads to more simplified expressions of the optimal transmit covariance matrix and the channel capacity. 
	
	First, we consider problem $(\ref{P0})$ for the MISO case with $N \ge 1$ and $M=1$. The effective channel is given by $\mathbf{h}(\mathbf{f}_{t},\mathbf{f}_{r})^H \!\!=\!\![h_{1,1}(\mathbf{f}_{t,1},\mathbf{f}_{r,1}), h_{1,2}(\mathbf{f}_{t,2},\mathbf{f}_{r,1}), \ldots, h_{1,N}(\mathbf{f}_{t,N},\mathbf{f}_{r,1})]$. The optimal transmit covariance matrix in the MISO case is obtained according to the maximum ratio transmission (MRT), which is given 
	\begin{align}
		\mathbf{Q}^{\star} = \frac{P_{\mathrm{max}}\mathbf{h}(\mathbf{f}_{t},\mathbf{f}_{r})\mathbf{h}(\mathbf{f}_{t},\mathbf{f}_{r})^H}{\Vert \mathbf{h}(\mathbf{f}_{t},\mathbf{f}_{r}) \Vert_2^2}.
	\end{align}
	The MISO channel capacity is expressed as
	\begin{align}
		C_{\mathrm{MISO}}(\mathbf{f}_t,\mathbf{f}_r) &= \log_2 \left( 1 + \frac{1}{\sigma^2} \mathbf{h}(\mathbf{f}_t,\mathbf{f}_r)^H \mathbf{Q} \mathbf{h}(\mathbf{f}_t,\mathbf{f}_r) \right) \nonumber \\
		&= \log_2 \left( 1 + \frac{P_{\mathrm{max}}}{\sigma^2} \Vert \mathbf{h}(\mathbf{f}_t,\mathbf{f}_r)\Vert_2^2 \right).
	\end{align}
	Thus, problem $(\ref{P0})$ is converted to the equivalent problem to maximize total channel power as
	\begin{subequations}
		\label{P3-2}
		\begin{eqnarray}
			\label{P3-2-0}
			&\!\!\!\!\!\!\! \max  \limits_{\mathbf{f}_t, \mathbf{f}_r}  
			&\!\!\!\! \Vert \mathbf{h}(\mathbf{f}_t, \mathbf{f}_r)\Vert_2^2 \\
			\label{P3-2-1}
			&\!\!\!\!\!\!\! \mathrm{s.t.}  &\!\!\!\! \eqref{P0-2},\eqref{P0-3},\eqref{P0-4}.
		\end{eqnarray}
	\end{subequations}
	Given $\{\mathbf{f}_{t,n}\}_{n=1}^N$, the objective function of problem $(\ref{P3-2})$ with respect to $\mathbf{f}_{r,1}$ can be formulated as
	\begin{align}
		\label{r_m}
		\mathbf{h}(\mathbf{f}_t, \mathbf{f}_r) =[\bar{f}_{1,1}, \bar{f}_{1,2}, \ldots, \bar{f}_{1,N}],
	\end{align}
	where
	\begin{align}
		&\bar{a}_{1,n}=a_{1,n}{{f}}_{t,1,n}, \bar{b}_{1,n,d}=b_{1,n,d}{{f}}_{t,n,d},
		\\
		&\bar{f}_{1,n}=\bar{a}_{1,n}{{f}}_{r,1,n} + \sum \nolimits_{d=1}^D \bar{b}_{1,n,d}{{f}}_{r,d,1}.
	\end{align}
	Then, the subproblem related to $\mathbf{f}_{r,1}$ can be expressed as 
	\begin{subequations}
		\label{P3-3}
		\begin{eqnarray}
			\label{P3-3-0}
			&\!\!\!\!\!\!\! \max  \limits_{\mathbf{f}_{r,1}}  
			&\!\!\!\! \mathbf{h}(\mathbf{f}_t, \mathbf{f}_r)^H\mathbf{h}(\mathbf{f}_t, \mathbf{f}_r) \\
			\label{P3-3-1}
			&\!\!\!\!\!\!\! \mathrm{s.t.}  &\!\!\!\! \Vert \mathbf{f}_{r,1} \Vert_2 = 1, \\
			\label{P3-3-2}
			&&\!\!\!\! \cos(\theta_{\max}) \le \mathbf{f}_{r,1}^T\mathbf{e}_z \le 1.
		\end{eqnarray}
	\end{subequations}
	Problem $(\ref{P3-3})$ can be solved by the proposed Algorithm \ref{alg:fw-ra}.
	
	Given $\{\mathbf{f}_{t,i}\}_{i=1, i \neq n}^N$ and $\mathbf{f}_{r,1}$, the objective function of problem $(\ref{P3-2})$ with respect to $\mathbf{f}_{t,n}$ can be formulated as
	\begin{subequations}
		\label{P3-4}
		\begin{eqnarray}
			\label{P3-4-0}
			&\!\!\!\!\!\!\! \max  \limits_{\mathbf{f}_{t,n}}  
			&\!\!\!\! \vert \tilde{f}_{1,n} \vert^2 \\
			\label{P3-4-1}
			&\!\!\!\!\!\!\! \mathrm{s.t.}  &\!\!\!\! \Vert \mathbf{f}_{t,n} \Vert_2 = 1, \\
			\label{P3-4-2}
			&&\!\!\!\! \cos(\theta_{\max}) \le \mathbf{f}_{t,n}^T\mathbf{e}_z \le 1,
		\end{eqnarray}
	\end{subequations}
	where
	\begin{align}
		&\tilde{a}_{1,n}=a_{1,n}{{f}}_{r,1,n}, \tilde{b}_{1,n,d}=b_{1,n,d}{{f}}_{r,d,1},
		\\
		&\tilde{f}_{1,n}=\tilde{a}_{1,n}{{f}}_{t,1,n} + \sum \nolimits_{d=1}^D \tilde{b}_{1,n,d}{{f}}_{t,n,d}.
	\end{align}
	Problem $(\ref{P3-4})$ can be solved by the proposed Algorithm \ref{alg:fw-ra}.
	
	We consider the special case with $p=1$ and $D=0$. For each $\mathbf{f}_{t,n}$ with fixed $\mathbf{f}_{r,1}$ and $\{\mathbf{f}_{t,n}\}_{i=1, i \neq n}^N$, we have
	\[
	\tilde f_{1,n}
	=
	a_{1,n} f_{r,1,n}
	(\mathbf R_t\mathbf f_{t,n})^T
	\mathbf u_{1,n},
	\]
	where $
	\mathbf u_{1,n}
	=
	\frac{\mathbf r_1-\mathbf t_n}{r_{1,n}}
	\in\mathbb R^3$.
	Since \(a_{1,n}f_{r,1,n}\) is independent of \(\mathbf f_{t,n}\),
	maximizing \(|\tilde f_{1,n}|^2\) is equivalent to maximizing $
	\vert
	(\mathbf R_t\mathbf f_{t,n})^T\mathbf u_{1,n}
	\vert^2$. Considering \((\mathbf R_t\mathbf f_{t,n})^T\mathbf u_{1,n}\ge0\),
	maximizing the squared projection is equivalent to maximizing the
	projection.
	Then, the subproblem is given by
	\begin{subequations}
		\label{P3-6}
		\begin{eqnarray}
			\label{P3-6-0}
			&\!\!\!\!\!\!\! \max  \limits_{\mathbf{f}_{t,n}}  
			&\!\!\!\! (\mathbf{R}_t\mathbf{f}_{t,n})^\mathrm{T}\mathbf u_{1,n}, \\
			\label{P3-6-1}
			&\!\!\!\!\!\!\! \mathrm{s.t.}  &\!\!\!\! \Vert \mathbf{f}_{t,n} \Vert_2 = 1, \\
			\label{P3-6-2}
			&&\!\!\!\! \cos(\theta_{\max}) \le \mathbf{f}_{t,n}^T\mathbf{e}_z \le 1,
		\end{eqnarray}
	\end{subequations}
	Since $\mathbf{R}_t$ is an orthogonal matrix and $(\mathbf{R}_t\mathbf{f}_{t,n})^\mathrm{T}\mathbf u_{1,n} \ge 0$, we rewrite the objective function \eqref{P3-6-0} as
	\begin{equation}
		(\mathbf{R}_t \mathbf{f}_{t,n})^T \mathbf u_{1,n} = \mathbf{f}_{t,n}^T (\mathbf{R}_t^T \mathbf u_{1,n}).
	\end{equation}
	Let $\mathbf{d} = \mathbf{R}_t^T \mathbf u_{1,n} = [d_x, d_y, d_z]^T$ be the effective gradient vector. The problem reduces to maximizing the inner product $\mathbf{f}_{t,n}^T \mathbf{d}$ within the feasible region $\mathcal{C}$.
	
	The optimal solution $\mathbf{f}_{t,n}^*$ is found by projecting $\mathbf{d}$ onto the feasible set defined by the unit sphere and the zenith angle limit. We define the horizontal component of $\mathbf{d}$ as $\mathbf{d}_{xy} = [d_x, d_y, 0]^T$.
	If the normalized vector $\mathbf{d}/\|\mathbf{d}\|$ satisfies the zenith angle constraint ${d_z}/{\|\mathbf{d}\|} \geq \cos(\theta_{\max})$.
	Then, the optimal solution is $\mathbf{f}_{t,n}^{\ast} = {\mathbf{d}}/{\|\mathbf{d}\|}$.
	If ${d_z}/{\|\mathbf{d}\|} < \cos(\theta_{\max})$, the solution must lie on the boundary of the zenith angle constraint $\mathbf{f}_{t,n}^T \mathbf{e}_z = \cos(\theta_{\max})$. The optimal vector maintains the same azimuth as $\mathbf{d}$ but is constrained to the maximum allowed angle
	\begin{equation}
		\!\!\! \mathbf{f}_{t,n}^{\ast} \!\!=\!\! \left[ \sin(\theta_{\max}) {d_x}/{d_0}, \sin(\theta_{\max}) {d_y}/{d_0}, \cos(\theta_{\max}) \right]^T,
	\end{equation}
	where $d_0 = \sqrt{d_x^2 + d_y^2}$.
	Thus, the optimal solution $\mathbf{f}_{t,n}^*$ can be computed as
	\begin{align}
		\label{MISO-solution}
		\!\!\!\! \mathbf{f}_{t,n}^{\ast} \!= \!\Pi_{\mathcal{C}}(\mathbf{d}) \!\!= \!\!
		\begin{cases} 
			{\mathbf{d}}/{\|\mathbf{d}\|}, & \text{if } {d_z}/{\|\mathbf{d}\|} \geq \cos(\theta_{\max}), \\
			\hat{\mathbf{f}}_{t,n} / \|\hat{\mathbf{f}}_{t,n}\|, & \text{otherwise},
		\end{cases}
	\end{align}
	where $\hat{\mathbf{f}}_{t,n}$ is the vector adjusted to the zenith boundary. The angle offset is given by $\epsilon_{t,n}= [\theta_{z,n}^{t,\star} - \theta_{\max}]_{+}$, where $\cos(\theta_{z,n}^t) = {d_z}/{\sqrt{d_x^2+d_y^2+d_z^2}}$. The details are presented in Appendix~\ref{app1}.
	
	Next, we consider the SIMO system with $M \ge 1$ and $N=1$, where the effective channel vector is given by $\label{channel_h1}
	\mathbf{h}(\mathbf{f}_{t},\mathbf{f}_{r})^H =[h_{1,1}(\mathbf{f}_{t,1},\mathbf{f}_{r,1}), h_{2,1}(\mathbf{f}_{t,1},\mathbf{f}_{r,2}), \ldots, h_{M,1}(\mathbf{f}_{t,1},\mathbf{f}_{r,M})]$.
	In this case, the optimal transmit covariance matrix is set as $\mathbf{Q}^{\star}=P_{\mathrm{max}}$. The channel capacity is given by
	\begin{align}
		C_{\mathrm{SIMO}}(\mathbf{f}_r,\mathbf{f}_t) &= \log_2 \left( 1 + \frac{1}{\sigma^2} \mathbf{h}(\mathbf{f}_r,\mathbf{f}_t)^H \mathbf{Q} \mathbf{h}(\mathbf{f}_r,\mathbf{f}_t) \right) \nonumber \\
		&= \log_2 \left( 1 + \frac{P_{\mathrm{max}}}{\sigma^2} \Vert \mathbf{h}(\mathbf{f}_r,\mathbf{f}_t)\Vert_2^2 \right).
	\end{align}
	It is notable that the capacity of SIMO channel given above takes a similar form to that of the MISO channel, which can be optimized by applying the same method.
	
	\section{Simulation Results}
	\label{Simulation Results}
	In this section, simulation results are provided to assess the effectiveness of the proposed algorithm. In the simulation, we consider a MIMO system with $N=16$ transmit RAs and $M=16$ receive RAs, where $M_x=M_y = 4$ and $N_x=N_y=4$. The scatterer locations $\{\mathbf s_d\}_{d=1}^D$
	are independently generated within the rectangular region between the
	transmitter and receiver, and the scattering phases are independently
	drawn as $\chi_d\sim\mathcal U[0,2\pi)$. The radar cross sections are set
	as $\sigma_d=5$. The carrier frequency of the system is $f=3.5$ GHz, where the wavelength is $\lambda = 0.0857$ m. We consider a global Cartesian coordinate system. The transmit array centers are placed at the origin, i.e., $\mathbf{t}_0=[0,0,0]$ m. The receiver array is located at $\mathbf{r}_0=[6,6,30]$ m. 
	The noise power is set to $\sigma^2 = -80~\mathrm{dBm}$, the antenna separation is ${\lambda}/{2}$. The transmit power is set to $P_{\mathrm{max}}=10~\mathrm{dBm}$, and the maximum zenith angle allowed for RA boresight rotation is set as $\theta_{\max} = \pi/6$. The number of scatterers is given by $D=6$. For performance evaluation, we consider the following benchmarks:
	\begin{itemize}
		\item 
		\textbf{FOA}: The transmitter and receiver are equipped with fixed UPAs. 
		\item 
		\textbf{Strongest eigenchannel power maximization (SEPM)}: The simplified algorithm presented in Section \ref{Solution for low-SNR regime} tailored for the low-SNR regime.
		\item 
		\textbf{RFOA}: The receiver is equipped with a fixed UPA, while the transmitter employs $N$ RAs. Thus, we only need to optimize $\mathbf{f}_t$ and $\mathbf{Q}$ in the proposed algorithm.
		\item 
		\textbf{TFOA}: The transmitter is equipped with a fixed UPA, while the receiver employs $M$ RAs. Thus, we only need to optimize $\mathbf{f}_r$ and $\mathbf{Q}$ in the proposed algorithm.
		\item 
		\textbf{Random-orientation baseline}: In this scheme, we randomly generate $\mathbf{f}_t$ and $\mathbf{f}_r$ satisfying the maximum zenith angle constraints for $1000$ independent realizations. Then, we obtain the channel capacity by optimizing $\mathbf{Q}$ with given $\mathbf{f}_t$ and $\mathbf{f}_r$ for each realization, and select  $\mathbf{f}_t$ and $\mathbf{f}_r$ with the largest capacity.
		\item 
		\textbf{Isotropic-antenna baseline}: In this benchmark, the antenna elements in the transmitter and receiver are isotropic, i.e., $p=0$, and the radiation energy is evenly distributed in the front half-space of the antennas. Thus, RA orientation optimization is not performed, and 
		only the transmit covariance matrix is optimized by the proposed algorithm.
	\end{itemize}
	
	\subsection{Convergence Behavior of Proposed Algorithm}
	\begin{figure}[htbp]
		\centering
		\includegraphics[width=0.43 \textwidth]{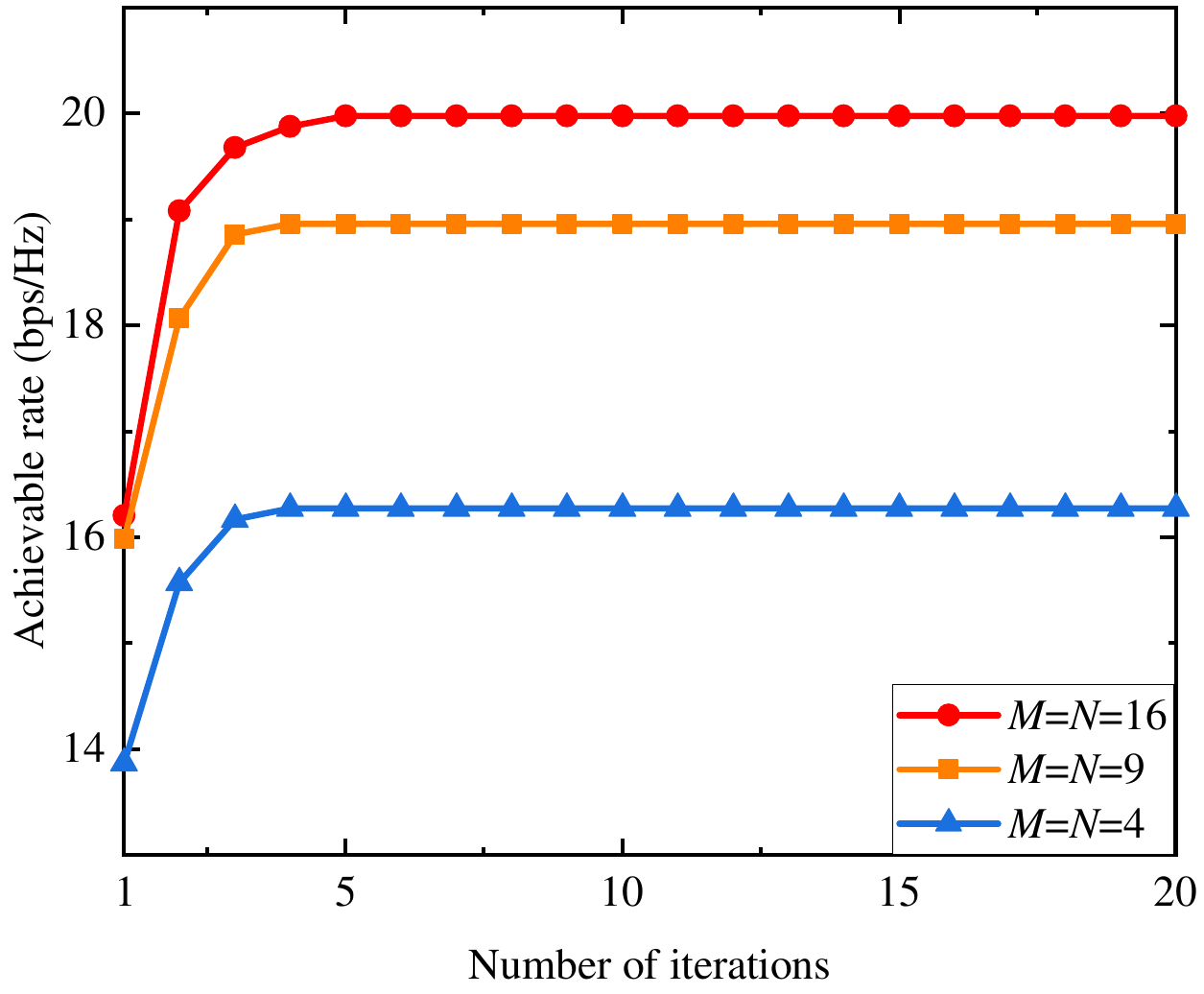}
		\caption{Achievable rate versus the number of iterations.}
		\label{Fig10}
	\end{figure}
	Fig.~\ref{Fig10} shows the convergence behavior of the proposed algorithm under different antenna numbers. It can be observed that the achievable rate monotonically increases with the number of iterations and quickly converges for all considered cases. Specifically, the proposed algorithm reaches a stable value within about $5$ iterations, which demonstrates its fast convergence and effectiveness.
	Moreover, the achievable rate increases with the number of antennas. In particular, the case with $M=N=16$ achieves the highest achievable rate, followed by $M=N=9$ and $M=N=4$. This is because a larger antenna array provides higher array gain and more spatial DoFs, which helps improve the effective MIMO channel power and spatial multiplexing capability.
	
	\subsection{Performance Analysis}
	
	\begin{figure}[t]
		\centering
		\begin{subfigure}{\columnwidth}
			\centering
			\includegraphics[width=0.9 \textwidth]{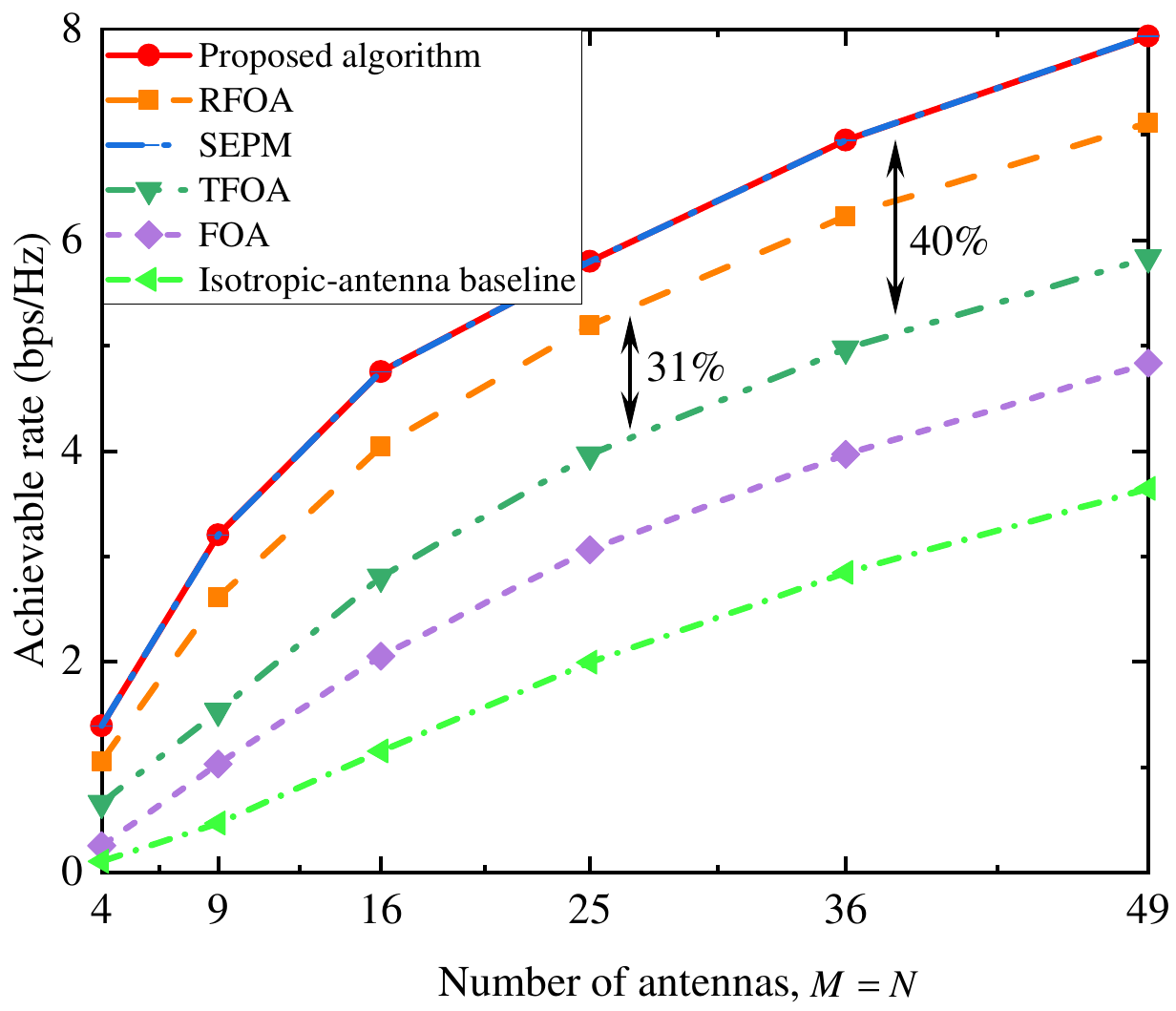}
			\caption{Achievable rate versus the number of antennas in low-SNR case, where $P_{\max}=-30~\mathrm{dBm}$.}
			\label{fig:sub-a}
		\end{subfigure}\\
		\begin{subfigure}{\columnwidth}
			\centering
			\includegraphics[width=0.9 \textwidth]{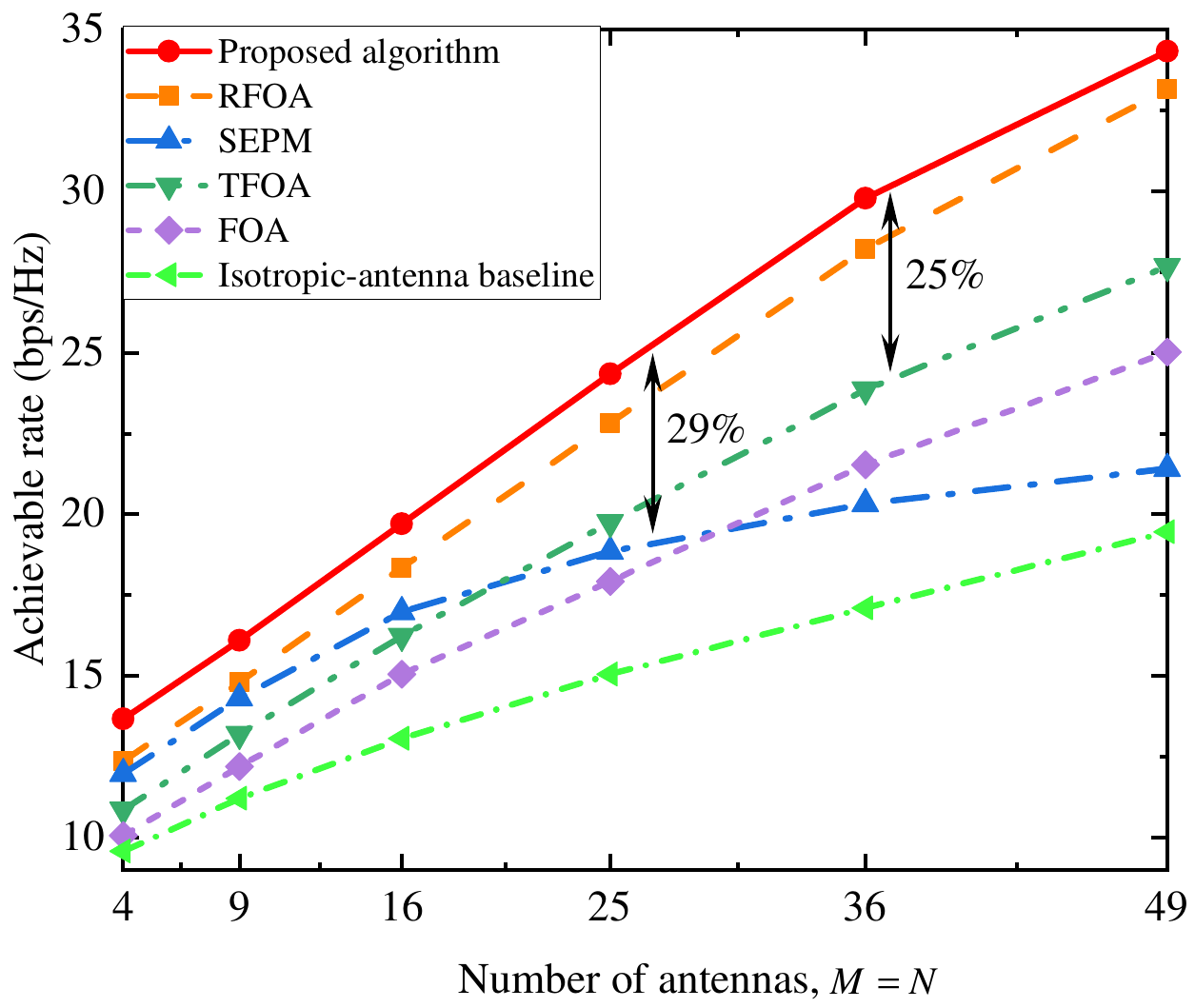}
			\caption{Achievable rate versus the number of antennas in high-SNR case, where $P_{\max}=10~\mathrm{dBm}$.}
			\label{fig:sub-b}
		\end{subfigure}
		
		\caption{Achievable rate versus the number of antennas.}
		\label{fig:total-figure}
		\vspace{-4mm}
	\end{figure}
	Fig.~\ref{fig:total-figure} shows the achievable rate versus the number of antennas $M$ and $N$ under the low-SNR and high-SNR regimes. It can be observed that the achievable rate of all schemes increases with the number of antennas. 
	In both cases, the proposed algorithm achieves the highest achievable rate, since it jointly optimizes the transmit and receive RA orientations and thus can better reshape the effective MIMO channel.
	As shown in Fig.~\ref{fig:sub-a}, in the low-SNR regime, 
	the performance gain of the proposed algorithm mainly comes from enhancing the dominant channel mode through transmit and receive orientation optimization. The SEPM scheme performs relatively close to the proposed algorithm in the low-SNR case, since it also focuses on strengthening the strongest eigenchannel.
	Furthermore, when $M=N=36$, the proposed algorithm achieves about $40\%$ gain over the TFOA scheme. This reveals the advantages from the joint optimization of both transmit-side and receiver-side antenna orientation. The TFOA scheme only optimizes the receive-side orientations, leading to limited channel reconfiguration capability.
	In addition, the RFOA scheme outperforms the TFOA scheme, where the RFOA achieves about $31\%$ gain over the TFOA scheme when $M=N=25$. This can be explained as follows. The RFOA scheme optimizes the transmit-side RA orientations, which can directly steer the radiated energy toward favorable propagation directions and reshape the effective MIMO channel from the source. As a result, both the desired channel power and the spatial eigenchannel structure can be improved before signal propagation. In contrast, TFOA only optimizes the receive-side RA orientations. Although it can enhance the reception of useful incoming paths, it cannot compensate for the power loss caused by mismatched transmit boresight directions. Therefore, transmit-side orientation optimization provides a stronger channel reconfiguration capability than receive-side-only optimization, leading to a higher achievable rate.

	Fig.~\ref{fig:sub-b} illustrates the achievable rate in the high-SNR regime.  As $M$ and $N$ increase, more spatial eigenchannels become available for 
	MIMO transmission, making spatial multiplexing increasingly important.  In this case, the proposed RA design improves
	the rate mainly by enhancing the effective gains of multiple spatial
	eigenchannels through orientation-domain alignment.
	Although the SEPM scheme can enhance the strongest eigenchannel, it tends to concentrate most of the channel power on the dominant mode and sacrifices the performance of other eigenchannels. By contrast, the proposed algorithm jointly optimizes the transmit and
	receive RA orientations together with the transmit covariance matrix,
	thereby increasing the effective singular values of the MIMO channel and
	achieving better high-SNR performance.
	Specifically, when $M=N=25$, the proposed algorithm achieves about $29\%$ gain over the SEPM scheme, while when $M=N=36$, it attains about $25\%$ gain over the TFOA scheme.

	\begin{figure}[t]
		\centering
		\includegraphics[width=0.43 \textwidth]{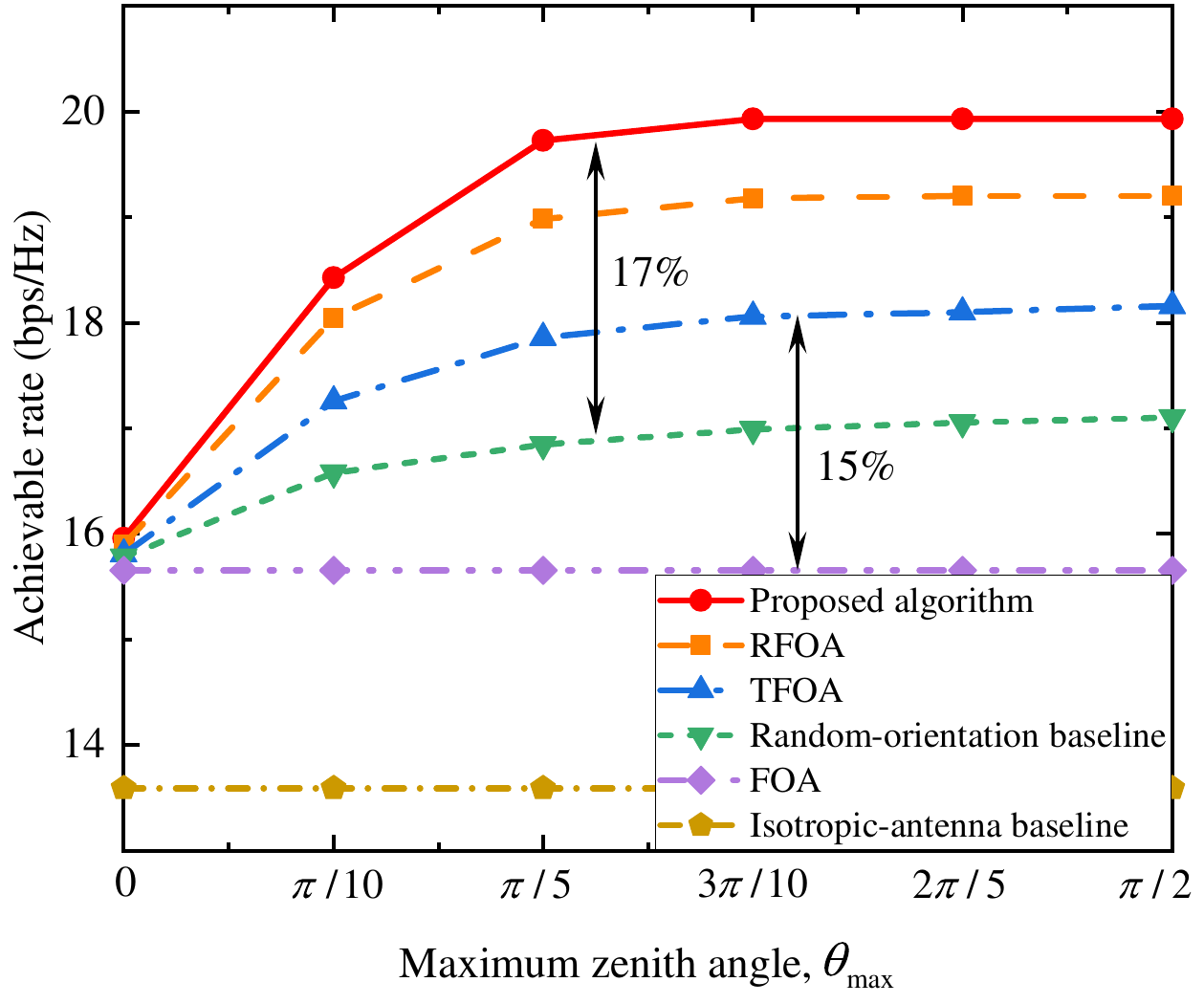}
		\caption{Achievable rate versus the maximum zenith angle.}
		\label{Fig5}
		\vspace{-4mm}
	\end{figure}
	
	Fig.~\ref{Fig5} shows the achievable rate versus the maximum zenith angle $\theta_{\max}$. As $\theta_{\max}$ increases, the achievable rate of the proposed algorithm increases and gradually converges. This is because a larger $\theta_{\max}$ enlarges the feasible orientation region of each RA, providing more orientation-domain DoFs to align the antenna boresights with favorable LoS and NLoS propagation directions.
	Specifically, when $\theta_{\max}$ increases from $0$ to $\pi/5$, the proposed algorithm improves rapidly and then becomes saturated. This indicates that the main useful propagation directions can already be covered when the allowed rotation range is sufficiently large. Further increasing $\theta_{\max}$ provides limited additional gain. 
	Moreover, the proposed algorithm achieves about $17\%$ gain over the random-orientation baseline, demonstrating the effectiveness of the proposed algorithm. Compared to FOA scheme, the TFOA scheme enjoys 15\% gain by optimizing the receive-side orientations, thereby improving the channel power and eigenchannel utilization.

	\begin{figure}[t]
		\centering
		\includegraphics[width=0.43 \textwidth]{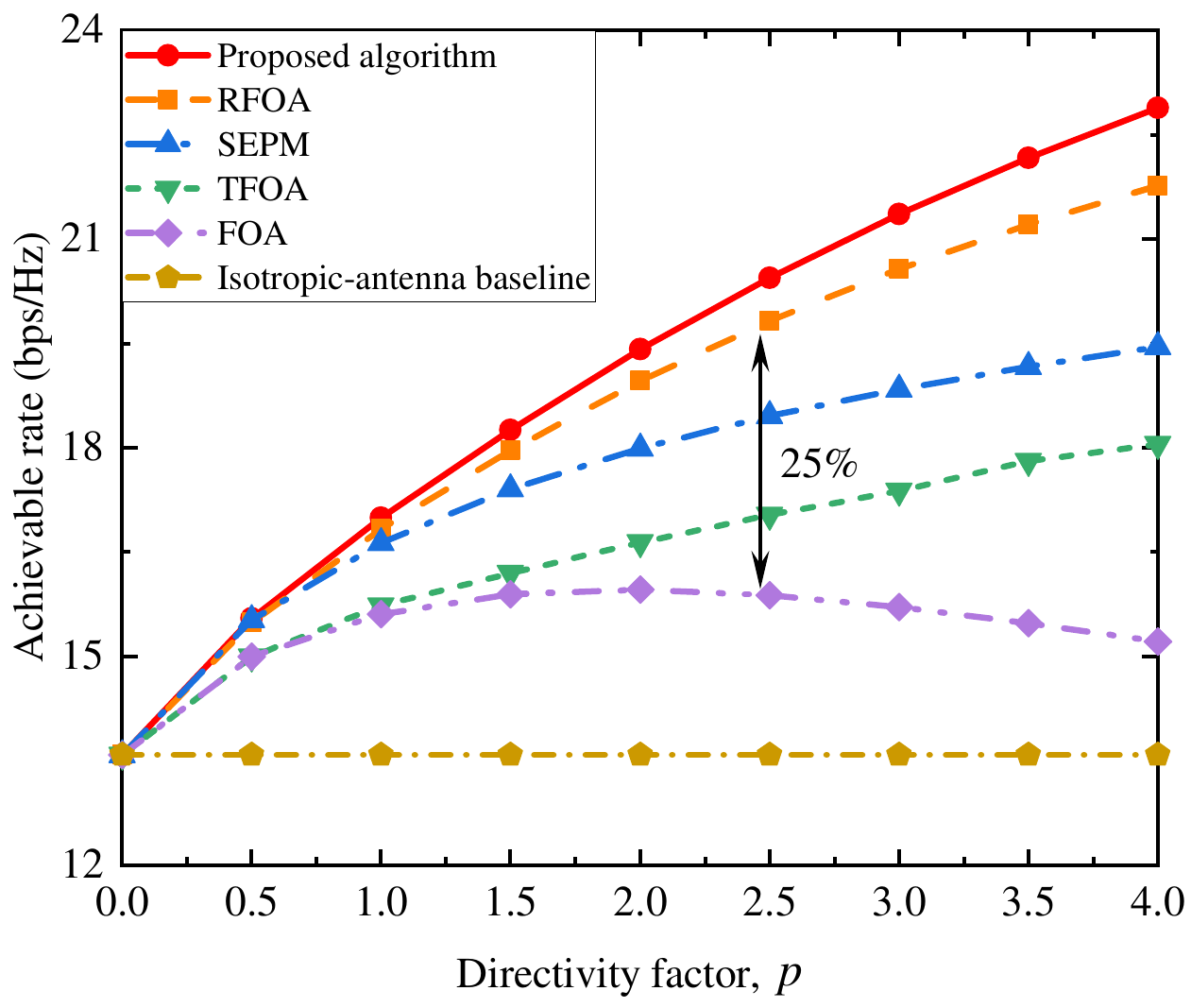}
		\caption{Achievable rate versus the directivity factor.}
		\label{Fig4}
		\vspace{-4mm}
	\end{figure}
	Fig.~\ref{Fig4} shows the achievable rate versus the directivity factor $p$. It can be observed that the achievable rate of the proposed algorithm increases with $p$, since a larger directivity factor leads to a narrower and stronger main lobe, which provides higher directional gain when the antenna boresight is properly aligned with favorable LoS and NLoS propagation paths. The proposed algorithm consistently achieves the highest achievable rate over the whole range of $p$, which demonstrates the effectiveness of jointly optimizing the transmit covariance matrix and the transmit/receive RA orientations.
	Compared with the FOA baseline, the RFOA scheme achieves about $25\%$ performance gain at $p=2.5$. This is reasonable because FOA employs fixed antenna orientations and thus cannot adapt the boresight directions to the propagation environment. 		
	In addition, the gap between the proposed algorithm and SEPM gradually becomes larger as $p$ increases. This can be explained as follows. SEPM scheme mainly aims to enhance the strongest eigenchannel, which is effective when the directivity gain is relatively limited. However, as $p$ becomes larger, the channel capacity depends not only on strengthening the dominant eigenchannel but also on maintaining a favorable balance among multiple eigenchannels. SEPM tends to concentrate most of the channel power on the strongest mode, which sacrifices the gains of the remaining eigenchannels and hence limits the spatial multiplexing performance. In contrast, the proposed algorithm jointly optimizes the RA orientations and the transmit covariance matrix, so that the increased directional gain can be more effectively exploited while preserving a more balanced eigenvalue distribution. Therefore, the performance advantage of the proposed algorithm over SEPM becomes more pronounced as $p$ increases.
	Moreover, the FOA curve slightly decreases in the large-$p$ region. This is because, under fixed antenna orientations, a more directional radiation pattern also makes the system more sensitive to boresight mismatch. As a result, part of the useful signal energy cannot be effectively captured when the main lobe deviates from the dominant propagation directions. This further indicates that the benefit of high antenna directivity can only be fully realized when the antenna orientations are properly optimized.
	
	\begin{figure}[t]
		\centering
		\includegraphics[width=0.43 \textwidth]{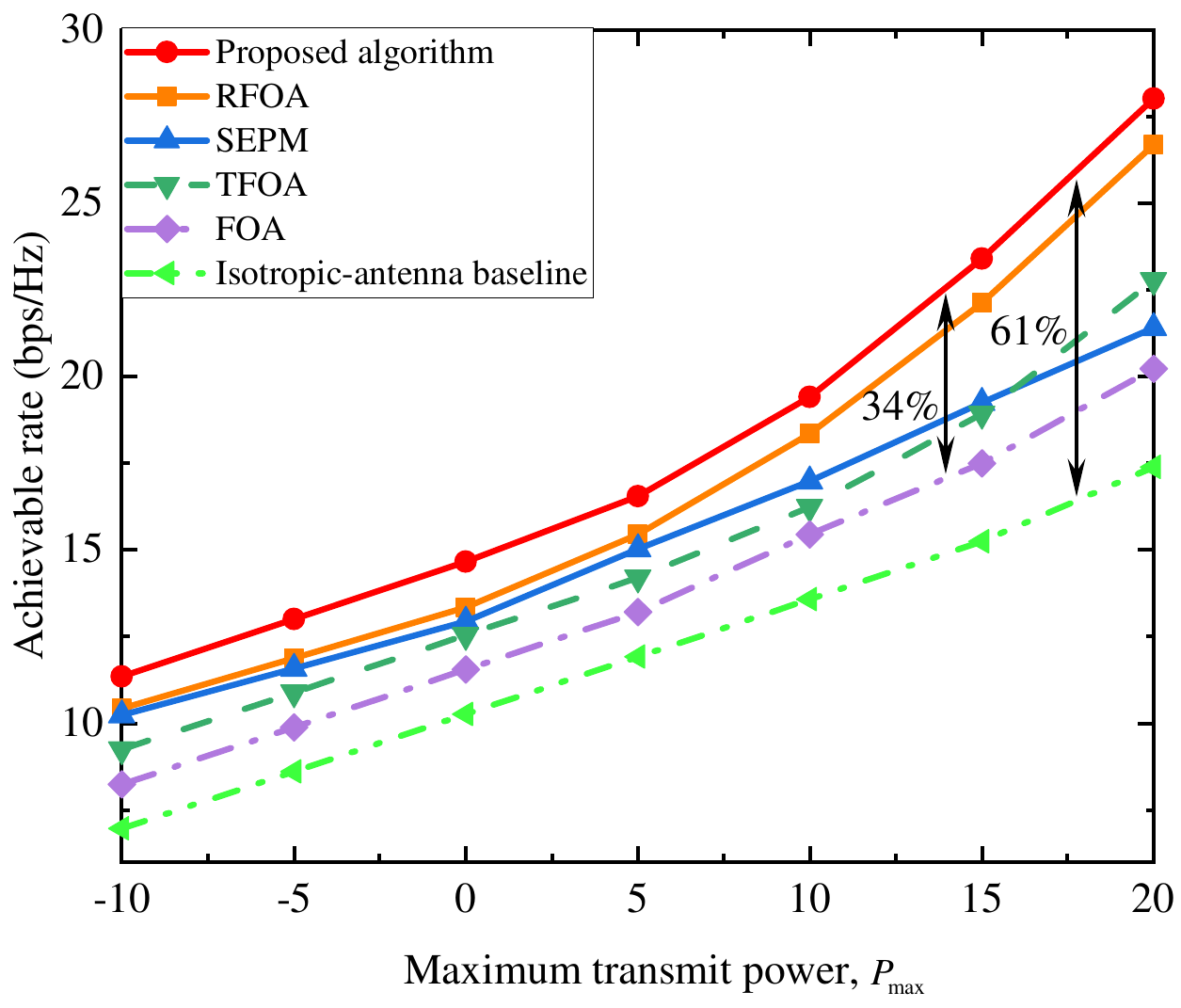}
		\caption{Achievable rate versus the maximum transmit power.}
		\label{Fig6}
		\vspace{-4mm}
	\end{figure}
	
	Fig.~\ref{Fig6} shows the achievable rate versus the maximum transmit power $P_{\max}$. It can be observed that the achievable rate of all considered schemes monotonically increases with $P_{\max}$. This is because a larger transmit power improves the received SNR, thereby enhancing the effective channel capacity.		
	Among all schemes, the proposed algorithm consistently achieves the highest achievable rate over the entire transmit-power range. This performance gain comes from the joint optimization of the transmit covariance matrix and the transmit/receive RA orientations, which enables more effective channel reconfiguration and better exploitation of both channel power enhancement and spatial multiplexing capability.
	In particular, at $P_{\max}=20$ dBm, the proposed algorithm achieves about $34\%$ gain over the FOA scheme and about $61\%$ gain over the isotropic-antenna baseline. 		
	Moreover, the RFOA scheme outperforms the TFOA scheme over the whole transmit-power range, indicating that optimizing the transmit-side RA orientations brings a stronger performance improvement than optimizing only the receive-side orientations.
	The SEPM scheme performs better than FOA in the low-power regime, but its gap from the proposed algorithm gradually widens as $P_{\max}$ increases, since SEPM mainly strengthens the strongest eigenchannel and is less effective in maintaining a favorable eigenvalue distribution for multi-stream transmission at high SNR.
	
	\vspace{-2mm}
	\section{Conclusion}
	\label{Conclusion}
	\vspace{-2mm}
	In this paper, we investigated an RA-aided MIMO communication system by adjusting the transmit and receive boresight directions to reshape the effective MIMO channel. We first established an orientation-dependent channel model, where the directional antenna gains were explicitly determined by the angular offsets between RA orientations and propagation paths.
	Based on the developed model, we formulated a capacity maximization problem by jointly optimizing the transmit covariance matrix and the transmit/receive RA orientations under practical spherical-cap constraints. To tackle the resulting non-convex problem, we proposed an AO framework, where the transmit covariance matrix was updated by eigenmode transmission and water-filling, and each RA orientation was optimized individually via a Riemannian Frank-Wolfe method. We further investigated the low-SNR regime and discussed MISO/SIMO special cases, which provided additional insights into the roles of dominant eigenchannel enhancement and simplified orientation design.
	Simulation results confirmed the advantages of the proposed RA-aided design. Compared with fixed-orientation antenna scheme, the proposed method achieved evident capacity improvements by enhancing the effective channel power and strengthening the useful singular modes of the MIMO channel. These results demonstrated that RA-aided MIMO offers a promising approach for channel reconfiguration and capacity enhancement in future wireless communication systems.
	
	\appendix
	
	\section{Derivation for obtaining (\ref{MISO-solution})} \label{app1}
	
	When the unconstrained solution is infeasible, the optimal solution must lie on the boundary of the spherical-cap constraint $\mathbf{f}^T \mathbf{e}_z = \cos\theta_{\max}$. We obtain
	\begin{subequations}
		\label{P9-1}
		\begin{eqnarray}
			\label{P9-1-0}
			&\!\!\!\!\!\!\! \max  \limits_{\mathbf{f}}  
			&\!\!\!\! \mathbf{f}^T \mathbf{d} \\
			\label{P9-1-1}
			&\!\!\!\!\!\!\! \mathrm{s.t.}  &\!\!\!\! \Vert {\mathbf{f}} \Vert_2 = 1, \\
			\label{P9-1-2}
			& &\!\!\!\!\mathbf{f}^T \mathbf{e}_z = c,
		\end{eqnarray}
	\end{subequations}
	where $c = \cos\theta_{\max}$.	We define the Lagrangian function
	\begin{equation}
		\mathcal{L}(\mathbf{f}, \lambda, \mu) = \mathbf{f}^T \mathbf{d} - \lambda (\mathbf{f}^T \mathbf{f} - 1) - \mu (\mathbf{f}^T \mathbf{e}_z - c),
	\end{equation}
	where $\lambda, \mu \in \mathbb{R}$ are Lagrange multipliers. Let $\nabla_{\mathbf{f}} \mathcal{L} = \mathbf{d} - 2\lambda \mathbf{f} - \mu \mathbf{e}_z = 0$. Then, we have
	\begin{equation}
		\mathbf{f} = {1}/{2\lambda} (\mathbf{d} - \mu \mathbf{e}_z). \label{eq:f_expr}
	\end{equation}
	Substituting \eqref{eq:f_expr} into the constraints \eqref{P9-1-1} and \eqref{P9-1-2}, we have
	\begin{align}
		&\frac{1}{2\lambda} (\mathbf{d}^T \mathbf{e}_z - \mu) = c \quad \Rightarrow \quad \mathbf{d}^T \mathbf{e}_z - \mu = 2\lambda c, \label{eq:mu_expr}\\
		&\frac{1}{4\lambda^2} \Vert {\mathbf{d} - \mu \mathbf{e}_z} \Vert^2 = 1 \quad \Rightarrow \quad \Vert{\mathbf{d} - \mu \mathbf{e}_z} \Vert^2 = 4\lambda^2. \label{eq:norm_eq}
	\end{align}
	
	Let $d_z = \mathbf{d}^T \mathbf{e}_z$ and $\mathbf{d}_{xy} = \mathbf{d} - d_z \mathbf{e}_z = (d_x, d_y, 0)^T$. Then $\Vert {\mathbf{d}_{xy}}\Vert^2 = \Vert{\mathbf{d}}\Vert^2 - d_z^2$.
	From \eqref{eq:mu_expr}, $\mu = d_z - 2\lambda c$. Substituting into \eqref{eq:norm_eq}, we have
	\begin{align}
		& \Vert{\mathbf{d}}\Vert^2 - 2(d_z - 2\lambda c)d_z + (d_z - 2\lambda c)^2 = 4\lambda^2, \\
		& \Vert{\mathbf{d}}\Vert^2 - 2d_z^2 + 4\lambda c d_z + d_z^2 - 4\lambda c d_z + 4\lambda^2 c^2 = 4\lambda^2, \\
		& \Vert{\mathbf{d}}\Vert^2 - d_z^2 + 4\lambda^2 c^2 = 4\lambda^2.
	\end{align}
	Thus, we have $\Vert{\mathbf{d}_{xy}}\Vert^2 = 4\lambda^2(1 - c^2)$.
	Let $s = \sqrt{1 - c^2}$. We obtain
	\begin{align}
		&\Vert{\mathbf{d}_{xy}}\Vert = 2\lambda s \quad \Rightarrow \quad 2\lambda = \frac{\Vert{\mathbf{d}_{xy}}\Vert}{s}. \label{eq:lambda}
	\end{align}
	\begin{align}
		&\mu = d_z - 2\lambda c = d_z - \frac{\Vert{\mathbf{d}_{xy}}\Vert c}{s}.
		\label{eq:mu_final}
	\end{align}
	Substituting \eqref{eq:lambda} and \eqref{eq:mu_final} into \eqref{eq:f_expr}, we have
	\begin{align}
		\mathbf{f} = \frac{1}{2\lambda} (\mathbf{d} - \mu \mathbf{e}_z) 
		&= \frac{s}{\Vert{\mathbf{d}_{xy}}\Vert} \left( \mathbf{d} - \left( d_z - \frac{\Vert{\mathbf{d}_{xy}}\Vert c}{s} \right) \mathbf{e}_z \right) \nonumber \\
		&= \frac{s}{\Vert{\mathbf{d}_{xy}}\Vert} (\mathbf{d} - d_z \mathbf{e}_z) + c \mathbf{e}_z \nonumber \\
		&= c \mathbf{e}_z + \frac{s}{\Vert{\mathbf{d}_{xy}}\Vert} \mathbf{d}_{xy}.
	\end{align}
	The above expression is valid for $\|\mathbf d_{xy}\|>0$. When $\|\mathbf d_{xy}\|=0$, i.e., $\mathbf d$ is collinear with $\mathbf e_z$, the boundary solution is not unique, and any feasible vector on the cone boundary with zenith angle $\theta_{\max}$ attains the same value.
	
	\vspace{-3mm}
	\bibliographystyle{IEEEtran}
	\begin{spacing}{0.95}
		\bibliography{thesis}

@ARTICLE{Saad6G,
	author={Saad, Walid and Bennis, Mehdi and Chen, Mingzhe},
	journal={IEEE Netw.}, 
	title={A Vision of {6G} Wireless Systems: Applications, Trends, Technologies, and Open Research Problems}, 
	year={Oct. 2019},
	volume={34},
	number={3},
	pages={134-142},
	doi={10.1109/MNET.001.1900287}}

@ARTICLE{Chen2022IRS,
	author={Chen, Guangji and Wu, Qingqing and He, Chong and Chen, Wen and Tang, Jie and Jin, Shi},
	journal={IEEE Trans. Wireless Commun.}, 
	title={Active {IRS} Aided Multiple Access for Energy-Constrained {IoT} Systems}, 
	year={Sept. 2022},
	volume={22},
	number={3},
	pages={1677-1694},
	doi={10.1109/TWC.2022.3206332}}

@ARTICLE{Zheng2026,
	author={Zheng, Ailing and Qingqing Wu and Ziyuan Zheng and Qiaoyan Peng and Yanze Zhu and Honghao Wang and Wen Chen and Guoying Zhang.},
	journal={arXiv preprint arXiv: 2601.04862}, 
	title={Wireless Communication with Cross-Linked Rotatable Antenna Array: Architecture Design and Rotation Optimization}, 
	year={2026},
	volume={},
	number={},
	pages={},
	doi={https://arxiv.org/abs/2601.04862}}

@ARTICLE{InterdonatoMIMO2020,
	author={Interdonato, Giovanni and Karlsson, Marcus and Björnson, Emil and Larsson, Erik G.},
	journal={IEEE Trans. Wireless Commun.}, 
	title={Local Partial Zero-Forcing Precoding for Cell-Free Massive {MIMO}}, 
	year={Apr. 2020},
	volume={19},
	number={7},
	pages={4758-4774},
	doi={10.1109/TWC.2020.2987027}}

@ARTICLE{Zhang20196G,
	author={Zhang, Zhengquan and Xiao, Yue and Ma, Zheng and others},
	journal={IEEE Veh. Tech. Mag.}, 
	title={{6G} Wireless Networks: Vision, Requirements, Architecture, and Key Technologies}, 
	year={Jul. 2019},
	volume={14},
	number={3},
	pages={28-41},
	doi={10.1109/MVT.2019.2921208}}

@ARTICLE{Zhu2024MA,
	author={Zhu, Lipeng and Ma, Wenyan and Zhang, Rui},
	journal={IEEE Trans. Wireless Commun.}, 
	title={Modeling and Performance Analysis for Movable Antenna Enabled Wireless Communications}, 
	year={Nov. 2023},
	volume={23},
	number={6},
	pages={6234-6250},
	doi={10.1109/TWC.2023.3330887}}

@ARTICLE{Ma2024MIMO,
		author={Ma, Wenyan and Zhu, Lipeng and Zhang, Rui},
		journal={IEEE Trans. Wireless Commun.}, 
		title={{MIMO} Capacity Characterization for Movable Antenna Systems}, 
		year={Sept. 2023},
		volume={23},
		number={4},
		pages={3392-3407},
		doi={10.1109/TWC.2023.3307696}}

@ARTICLE{Kaushik2019,
	author={Kaushik, Aryan and Thompson, John and Vlachos, Evangelos and Tsinos, Christos and Chatzinotas, Symeon},
	journal={IEEE Trans. Green Commun. Netw.}, 
	title={Dynamic {RF} Chain Selection for Energy Efficient and Low Complexity Hybrid Beamforming in Millimeter Wave {MIMO} Systems}, 
	year={Jul. 2019},
	volume={3},
	number={4},
	pages={886-900},
	doi={10.1109/TGCN.2019.2931613}}

@ARTICLE{Teng2022,
	author={Teng, Yinglei and Zhao, Yangliu and Wei, Min and Liu, An and Lau, Vincent K. N.},
	journal={IEEE Trans. Wireless Commun.}, 
	title={Sparse Hybrid Precoding for Power Minimization With an Adaptive Antenna Structure in Massive {MIMO} Systems}, 
	year={Jan. 2022},
	volume={21},
	number={7},
	pages={5279-5292},
	doi={10.1109/TWC.2021.3139097}}

@ARTICLE{Chowdhury6G2020,
	author={Chowdhury, Mostafa Zaman and Shahjalal, Md. and Ahmed, Shakil and Jang, Yeong Min},
	journal={IEEE Open J. Commun. Soc.}, 
	title={{6G} Wireless Communication Systems: Applications, Requirements, Technologies, Challenges, and Research Directions}, 
	year={Jul. 2020},
	volume={1},
	number={},
	pages={957-975},
	doi={10.1109/OJCOMS.2020.3010270}}

@ARTICLE{LuMIMO2014,
	author={Lu, Lu and Li, Geoffrey Ye and Swindlehurst, A. Lee and Ashikhmin, Alexei and Zhang, Rui},
	journal={IEEE J. Sel. Topics Signal Process.}, 
	title={An Overview of Massive {MIMO}: Benefits and Challenges}, 
	year={Apr. 2014},
	volume={8},
	number={5},
	pages={742-758},
	doi={10.1109/JSTSP.2014.2317671}}

@ARTICLE{Peng2025,
			author={Peng, Xingxiang and Wu, Qingqing and Zheng, Ziyuan and Chen, Wen and Zhu, Yanze and Gao, Ying},
			journal={IEEE Trans. Wireless Commun.}, 
			title={Rotatable Antenna Enabled Spectrum Sharing: Joint Antenna Orientation and Beamforming Design}, 
			year={early access, Apr. 2026},
			volume={},
			number={},
			pages={},
			doi={10.1109/TWC.2026.3683870}}

@ARTICLE{Zhu2024CM,
	author={Zhu, Lipeng and Ma, Wenyan and Zhang, Rui},
	journal={IEEE Commun. Mag.}, 
	title={Movable Antennas for Wireless Communication: Opportunities and Challenges}, 
	year={Oct. 2023},
	volume={62},
	number={6},
	pages={114-120},
	doi={10.1109/MCOM.001.2300212}}

@ARTICLE{Wong2021,
	author={Wong, Kai-Kit and others},
	journal={IEEE Trans. Wireless Commun.}, 
	title={Fluid Antenna Systems}, 
	year={Nov. 2020},
	volume={20},
	number={3},
	pages={1950-1962},
	doi={10.1109/TWC.2020.3037595}}

@ARTICLE{New2024,
		author={New, Wee Kiat and Wong, Kai-Kit and Xu, Hao and Tong, Kin-Fai and Chae, Chan-Byoung},
		journal={IEEE Trans. Wireless Commun.}, 
		title={An Information-Theoretic Characterization of {MIMO-FAS}: Optimization, Diversity-Multiplexing Tradeoff and q-Outage Capacity}, 
		year={Oct. 2023},
		volume={23},
		number={6},
		pages={5541-5556},
		doi={10.1109/TWC.2023.3327063}}

@ARTICLE{Li2025,
		author={Li, Zhendong and Ba, Jianle and Su, Zhou and Peng, Haixia and Wang, Yuntao and Chen, Wen and Wu, Qingqing},
		journal={IEEE Trans. Wireless Commun.}, 
		title={Joint Discrete Antenna Positioning and Beamforming Optimization in Movable Antenna Enabled Full-Duplex {ISAC} Networks}, 
		year={Nov. 2025},
		volume={25},
		number={},
		pages={7220-7234},
		doi={10.1109/TWC.2025.3630154}}

@ARTICLE{Liu2025,
			author={Liu, Wenchao and others},
			journal={IEEE Wireless Commun. Lett.}, 
			title={{UAV}-Enabled Wireless Networks With Movable-Antenna Array: Flexible Beamforming and Trajectory Design}, 
			year={Aug. 2025},
			volume={14},
			number={3},
			pages={566-570},
			doi={10.1109/LWC.2024.3451246}}

@ARTICLE{Tang2025,
		author={Tang, Xiao-Wei and Shi, Yunmei and Huang, Yi and Wu, Qingqing},
		journal={IEEE Wireless Commun. Lett.}, 
		title={Joint Optimization of {UAV} Height and Antenna Configuration for {UAV}-Mounted Movable Antenna}, 
		year={Oct. 2025},
		volume={15},
		number={},
		pages={235-239},
		doi={10.1109/LWC.2025.3623464}}

@ARTICLE{Xiao2025NOMA,
	author={Xiao, Zhenyu and others},
	journal={IEEE Trans. Wireless Commun.}, 
	title={Movable Antenna Aided {NOMA}: Joint Antenna Positioning, Precoding, and Decoding Design}, 
	year={Sept. 2025},
	volume={25},
	number={},
	pages={4595-4612},
	doi={10.1109/TWC.2025.3612424}}

@ARTICLE{Shao2025-1,
	author={Shao, Xiaodan and Zhang, Rui},
	journal={IEEE Commun. Mag.}, 
	title={{6DMA} Enhanced Wireless Network with Flexible Antenna Position and Rotation: Opportunities and Challenges}, 
	year={Mar. 2025},
	volume={63},
	number={4},
	pages={121-128},
	doi={10.1109/MCOM.002.2400333}}

@ARTICLE{Shao2025-2,
	author={Shao, Xiaodan and others},
	journal={IEEE J. Sel. Areas Commun.}, 
	title={{6D} Movable Antenna Enhanced Wireless Network via Discrete Position and Rotation Optimization}, 
	year={Jan. 2025},
	volume={43},
	number={3},
	pages={674-687},
	doi={10.1109/JSAC.2025.3531571}}

@ARTICLE{Ren2025,
	author={Ren, Tianshi and others},
	journal={IEEE Wireless Commun. Lett.}, 
	title={{6-D} Movable Antenna Enhanced Interference Mitigation for Cellular-Connected {UAV} Communications}, 
	year={Mar. 2025},
	volume={14},
	number={6},
	pages={1618-1622},
	doi={10.1109/LWC.2025.3549502}}

@ARTICLE{Gao2025,
	author={Gao, Ying and Wu, Qingqing and Chen, Wen},
	journal={IEEE Trans. Wireless Commun.}, 
	title={Movable Antennas Enabled Wireless-Powered {NOMA}: Continuous and Discrete Positioning Designs}, 
	year={Nov. 2025},
	volume={25},
	number={},
	pages={7132-7147},
	doi={10.1109/TWC.2025.3629352}}

@ARTICLE{Liang2025,
		author={Dai, Liang and others},
		journal={IEEE Wireless Commun. Lett.}, 
		title={Rotatable Antenna-Enabled Secure Wireless Communication}, 
		year={Jul. 2025},
		volume={14},
		number={11},
		pages={3440-3444},
		doi={10.1109/LWC.2025.3593258}}

@ARTICLE{Zhou2025,
	author={Zhou, Chao and You, Changsheng and Zheng, Beixiong and Shao, Xiaodan and Zhang, Rui},
	journal={IEEE Wireless Commun. Lett.}, 
	title={Rotatable Antennas for Integrated Sensing and Communications}, 
	year={Jun. 2025},
	volume={14},
	number={9},
	pages={2838-2842},
	doi={10.1109/LWC.2025.3580889}}

@ARTICLE{Zhang2025RA,
		author={Zhang, Xuzhong and others},
		journal={IEEE Trans. Commun.}, 
		title={Rotatable Antenna Array Enabled {UAV mmWave} Massive {MIMO} Communication}, 
		year={Oct. 2025},
	    volume={74},
		number={},
		pages={1219-1236},
		doi={10.1109/TCOMM.2025.3622962}}

@ARTICLE{Zhou2024NOMA,
	author={Zhou, Yufeng and others},
	journal={IEEE Wireless Commun. Lett.}, 
	title={Movable Antenna Empowered Downlink {NOMA} Systems: Power Allocation and Antenna Position Optimization}, 
	year={Aug. 2024},
	volume={13},
	number={10},
	pages={2772-2776},
	doi={10.1109/LWC.2024.3445110}}

@ARTICLE{Zheng2025MA-4,
		author={ Zheng, Ziyuan and  Wu, Qingqing and  Zhu, Yanze and  Wang, Honghao and  Gao, Ying and Chen, Wen and Xiong, Jian  },
		journal={arXiv preprint arXiv: 2512.20987}, 
		title={Low-Altitude {ISAC} with Rotatable Active and Passive Arrays}, 
		year={2025},
		volume={},
		number={},
		pages={},
		doi={https://arxiv.org/abs/2512.20987}}

@ARTICLE{Qiansecure2026,
	author={Qian, Yanzhi and others},
	journal={IEEE Commun. Lett.}, 
	title={{6DMA}-Assisted Secure Wireless Communications}, 
	year={Mar. 2026},
	volume={30},
	number={},
	pages={1499-1503},
	doi={10.1109/LCOMM.2026.3676904}}

@ARTICLE{Guo2024MA,
			author={Guo, Yuan and Chen, Wen and Wu, Qingqing and Liu, Yang and Wu, Qiong and Wang, Kunlun and Li, Jun and Xu, Lexi},
			journal={IEEE Trans. Wireless Commun.}, 
			title={Movable Antenna Enhanced Networked Integrated Sensing and Communication System}, 
			year={Oct. 2025},
			volume={25},
			number={},
			pages={5555-5572},
			doi={10.1109/TWC.2025.3619214}}

@ARTICLE{Wang2025MA-IRS,
		author={ Wang, Honghao and   Wu,Qingqing and   Jiang,Yifan and  Zheng, Ziyuan and  Zhang, Ziheng and  Zhu, Yanze and  Gao,Ying and   Chen,Wen and  Liu,Guanghai and  Jamalipour, Abbas},
		journal={arXiv preprint arXiv: 2511.10310}, 
		title={Reconfigurable Airspace: Synergizing Movable Antenna and Intelligent Surface for Low-Altitude {ISAC} Networks}, 
		year={2024},
		volume={},
		number={},
		pages={},
		doi={https://arxiv.org/abs/2511.10310}}

@article{Jiang2025RA,
			title={Average Secrecy Capacity Maximization of Rotatable Antenna-Assisted Secure Communications},
			author={Jiang, Pengchuan and Li, Quanzhong  and Mai, Lifeng and Zhang, Qi},
			journal={arXiv preprint arXiv:2601.04862}, 
			year={2026}}

@ARTICLE{ChenUAV2026,
			author={Chen, Shiying and others},
			journal={IEEE Trans. Veh. Tech.}, 
			title={Rotatable Antenna Meets {UAV}: Towards Dual-Level Channel Reconfiguration Paradigm for {ISAC}}, 
			year={early access, Apr. 2026},
			volume={},
			number={},
			pages={},
			doi={10.1109/TVT.2026.3682292}}

@ARTICLE{XueRACSI,
	author={Xiong, Xue and others},
	journal={IEEE Wireless Commun. Lett.}, 
	title={Efficient Channel Estimation for Rotatable Antenna-Enabled Wireless Communication}, 
	year={Aug. 2025},
	volume={14},
	number={11},
	pages={3719-3723},
	doi={10.1109/LWC.2025.3601979}}
	\end{spacing}
	
\end{document}